# Angstrofluidics: walking to the limit

Yi You [1,2,#], Abdulghani Ismail[1,2,#], Gwang-Hyeon Nam[1,2], Solleti Goutham[1,2], Ashok Keerthi [2,3] and Boya Radha[1,2,*]

[1]Department of Physics and Astronomy, School of Natural Sciences, The University of Manchester, Manchester M13 9PL, United Kingdom

[2]National Graphene Institute, The University of Manchester, Manchester M13 9PL, United Kingdom

[3]Department of Chemistry, School of Natural Sciences, The University of Manchester, Manchester M13 9PL, United Kingdom

[#] These authors contributed equally.

*Correspondence to: radha.boya@manchester.ac.uk

## Abstract

Angstrom-scale fluidic channels are ubiquitous in nature, and play an important role in regulating cellular traffic, signaling, and responding to stimuli. Synthetic channels are now a reality with the emergence of several cutting-edge bottom-up and top-down fabrication methods. In particular, the use of atomically thin two dimensional (2D) materials and nanotubes as components to build fluidic conduits has pushed the limits of fabrication to the Angstrom-scale. Here, we provide an overview of the recent developments in the fabrication methods for nano- and angstrofluidic channels while categorizing them on the basis of dimensionality (0D pores, 1D tubes, 2D slits), along with the latest advances in measurement techniques. We discuss the ionic transport governed by various stimuli in these channels and draw comparison of ionic mobility, streaming and osmotic power, with varying pore sizes across all the dimensionalities. Towards the end of the review, we highlight the unique future opportunities in the development of smart ionic devices.

# 1. Introduction

The art of making confined spaces for studying fluidics and various physical/chemical phenomena has been practiced for over half a century. Especially after the development of microfabrication techniques for semiconductors and microfiltration processes, the domain of microfluidics flourished in the 1980's with applications in gas chromatography, chemical analysis and cell biology (1-3). The ability to control the volume of fluids and to miniaturize the sample sizes was seen as the main advantages of microfluidics. Nanofluidics, where at least one of the channel dimensions is below 100 nm, emerged as an independent field after approximately two decades. In nanofluidics, the fluid behavior deviates drastically from that observed in microfluidics, with surface effects dominating due to an increase in the surface-to-volume ratio.

With the advent of new materials and advances in nanofabrication, the nanofluidics field has steadily walked down the path toward angstrofluidics, pushing the limits of fluidics (**Figure 1**), somewhat akin to how advances in semiconductor chips, following Moore's law, led to the latest 1-nm transistors (4). The discoveries of nanomaterials such as carbon nanotubes (CNTs) (5) and graphene (6), and new fabrication methods (7-9) such as controlled etching, ion and electron- beam sculpting, nanomanipulation and van der Waals assembly (10) have provided researchers with atomic scale building blocks and reproducible methodologies to design nano- and angstrom (Å) channels on demand.

What makes angstrofluidics interesting is that it deals with molecular interactions in a confined space at the finest level, where the confinement is smaller than the range of van der Waals (~1 to 50 nm) and steric-hydration interactions (1 to 2 nm) (11). For instance, 30 water molecules can barely fit in in 0.8-nm wide, 10-nm long CNTs,  yet they conduct $2.3 \times 10^{10}$ water molecules per second (12). In a graphene slit-like channel with a height of < 4 Å, only one molecular layer of water can fit in with no neighbors above or below to form intermolecular hydrogen bonds. Such behavior solicits an analogy with biological channels such as aquaporins, which allow only one water molecule through their narrowest constriction at a given point and yet conduct ultra-fast flows [$10^9$ molecules per second (12)]. The molecular and ionic flows through artificial Å-channels are comparable and sometimes higher than those through biological channels. The protein channels' selectivity toward a specific molecule or ion and their response to stimuli forces (gating) can be remarkable; for example, the selectivity of potassium channels toward $Na^+$ is at least $10^4$:1 (13). Research into mimicking biological channels with artificial systems aims to better understand and (at least partly) replicate biological channels' exotic functions, including active ion pumping and neurological signaling, and their mechanosensitive and piezosensitive nature (14). The electrochemical characteristics of ions in artificial Å-channels change due to the small dimensions of the channels, which interfere with the ion's own space and induce direct modifications of its behavior. The classical notions of molecular and ionic interactions in the bulk are no longer valid at such length scales; this has been researched using theoretical simulations for more than a decade, despite still being science fiction experimentally.

Together with advances in fabrication methods, the development of measurement techniques that describe fluid behavior deviating from the continuum is another critical reason for the progress of this field. Nano- and angstrofluidic systems with precise dimensions have shown diverse new phenomena such as fast water and gas transport (15, 16), non-linear electrokinetic transport (17), higher viscosity (18), and higher proton mobility under confinement (19) and anomalies such as lower water dielectric constant (20) compared with bulk. Therefore, it is important to ruminate on the key factors enabling remarkable advances in this domain for further new findings based on rational designs.

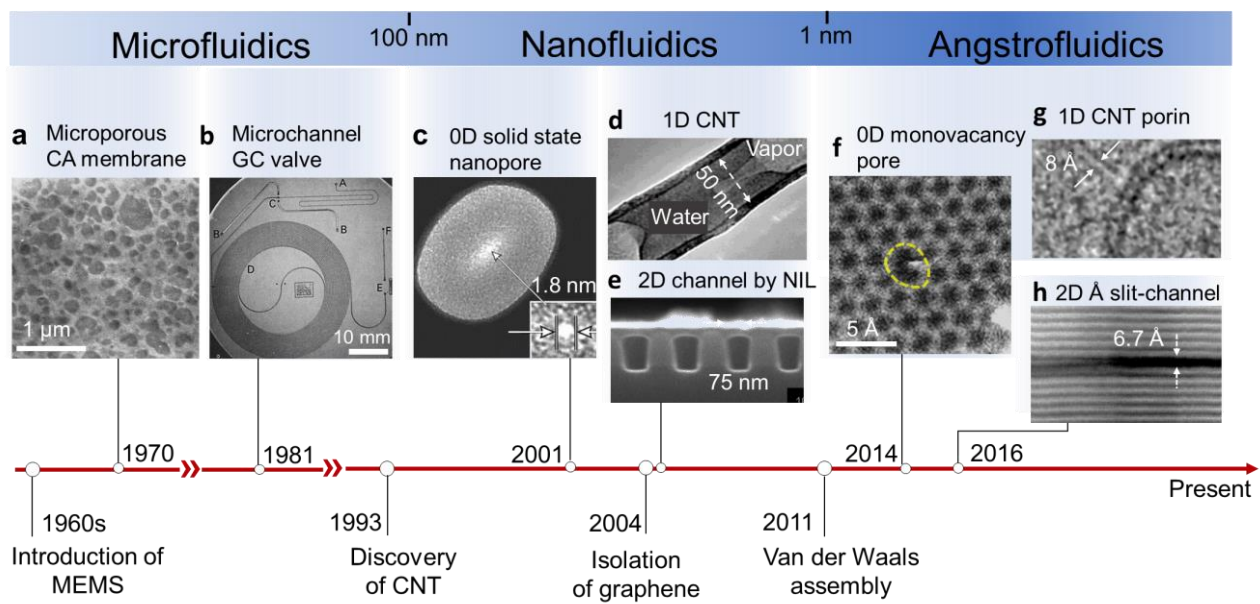

**Figure 1 Timeline of microfluidics, nanofluidics, and angstrofluidics.** This roadmap shows major milestones and advances in techniques and materials from the micro-electromechanical system (MEMS) to angstrofluidics. Within each fluidic field, selected examples of 0D, 1D, and 2D porous structures for mass transport are highlighted (a) Microporous cellulose acetate (CA) membrane. (b) Microchannel gas chromatograph (GC) valve. (c) 0D solid state silicon nitride nanopore. (d) 1D carbon nanotube (CNT). (e) 2D $SiO_2/Si$ channels by nanoimprint lithography (NIL). (f) 0D monovacancy pore on graphene membrane. (g) 1D CNT porin-liposome. (h) Angstrom slit made from $MoS_2$ layers. Panel *a* adapted with permission from Elsevier, copyright 1970, Reference 21. Panel *b* adapted with permission from Reference 22; copyright 1981, Elsevier. Panel *c* adapted with permission from Reference 23; copyright 2001, Springer Nature. Panel *d* adapted with permission from Reference 24; copyright 2004, American Chemical Society. Panel *e* adapted with permission from Reference 25; copyright 2004, American Chemical Society. Panel *f* adapted with permission from Reference 26; copyright 2014, American Chemical Society. Panel *g* adapted with permission from Reference 27; copyright 2016, Springer Nature. Panel *h* adapted with permission from Reference 16; copyright 2018, Springer Nature.

Here we review nanofluidics' transition to angstrofluidics with a focus on ionic transport in such systems and several projected applications. Excellent reviews that tackle the knowledge gaps in understanding sub-10 nm fluidics (28), mechanisms and governing equations of fluid flows (8, 9), and application of nanofluidic systems (7, 29–32) are available in the literature. Our current review is organized as follows: We start by describing fabrication advances for producing angstrom-scale channels, followed by several of the latest measurement techniques and their limitations. We then discuss ionic transport in response to different stimuli (electrical, pressure, osmotic, multi-stimuli) and the most recent developments in smart devices. We conclude with a perspective on the current applications of this domain and how we see its future.

## 2. Fabrication methods of nano/angstrofluidic channels

Based on their dimensions, fluidic channels can be divided into three categories: quasi-zero-dimensional (0D) pores, one-dimensional (1D) tubes, and two-dimensional (2D) slits (7). Pores have diameters of similar dimensions to their lengths ($d \approx l$) while tubes have a diameter much smaller than their length ($d \ll l$). Slits are composed of systems that have only one dimension (i.e., height) smaller than the other dimensions (i.e., length and width) ($h \ll w \leq l$). In this section, we provide a list of representative fabrication methods for nano- and angstrofluidic channels of all dimensionalities, grouped into two broad categories as either top-down or bottom-up methods.

## 2.1 Top-Down methods

Top-down methods often involve creating channels on a nonporous material with a variety of advanced techniques such as focused ion or electron beam, lithography, or etching (33). With top-down approaches, channels with reproducible and well-controlled dimensions can be achieved with high precision to make single- or multi-channel devices.

### 2.1.1. 0D pores

Solid-state nanopores are primary examples of 0D nanofluidic systems that have been studied extensively for biomolecular sensing and molecular separation (34-36). With the aid of electron beam lithography and ion etching techniques, one can synthesize free-standing solid-state pores with sizes ranging from a few nanometers to micrometers (35, 36). There have been a number of attempts to manipulate the pore size below the 10 nm regime, with the aim to sequence DNA. Danelon et al. (37) showed an electron beam-induced deposition of silica to shrink a pre-formed pore in a silicon nitride (SiN) membrane from ~ 50 nm to less than 10 nm in diameter. In another study, Storm et al. (34) used a high energy electron beam (e-beam) to soften the silica layer surrounding the pore, leading to a reduction of pore size from 20 nm to 2 nm (**Figure 2a**). More recently, the advent of 2D materials has led to an upsurge of research on atomically thin membranes that hold the potential for sequencing with single-nucleotide resolution and single-ion transport. Using advanced drilling techniques, angstrom (Å) sized pores can be created in such atomically thin membranes. Feng et al. (38) applied electrochemical etching to make Å-pores of ~ 0.6 nm on a single layer of $MoS_2$ membrane. Likewise, Thiruraman et al. (39) fabricated angstrom-pores created by well-controlled $Ga^+$ ion beam irradiation, presented as atomic vacancy defects, with a median diameter ranging from 0.2 nm to 0.5 nm on a single layer of $MoS_2$ membrane(**Figure 2b**).

### 2.1.2. 1D tubes

Chemical etching of conical channels and incorporation of nanotubes in membranes are common approaches for fabricating 1D nanofluidic systems. For instance, Zhang et al. (40) demonstrated a method of ion track etching followed by chemical etching to make a single conical channel of 6-nm diameter at the tip in a polyethylene terephthalate (PET) membrane. Alternatively, the incorporation of CNTs into SiN membranes can be realized in a precise fashion using a scanning electron microscopy (SEM)-assisted nanomanipulation method (41). Briefly, a single nanotube was inserted into a pre-drilled hole on a SiN membranes and glued by an e-beam-induced deposition of hydrocarbons (**Figure 2c**) (41). Using this method, membranes containing a single boron nitride nanotube (BNNT) with inner diameter of 20 nm (41) and a CNT with inner diameter of ~4 nm (42) were demonstrated.

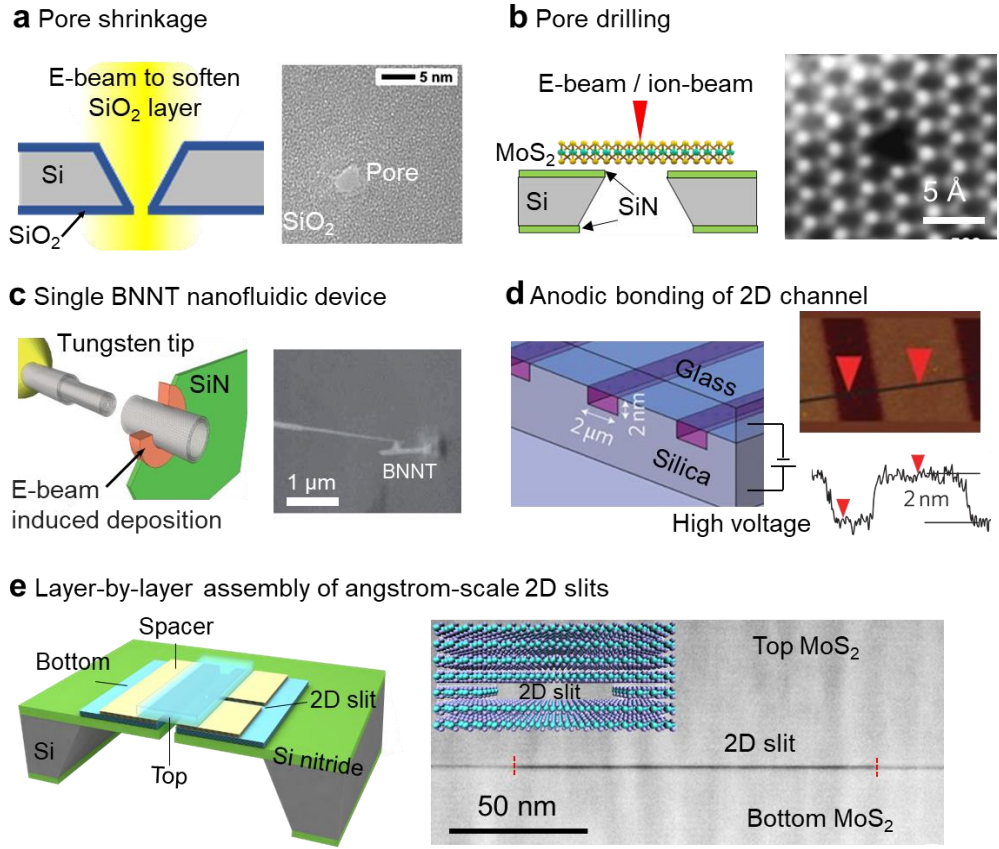

**Figure 2 State-of-the-art top-down fabrication methods of nano- and angstrofluidic systems.** (*a*) A 3-nm single pore in a silicon-based membrane made using electron-beam (e-beam) induced pore shrinkage. The final pore geometry and dimension is shown in the transmission electron microscopy image. (*b*) An Å-pore in a free-standing monolayer $MoS_2$ membrane drilled by a focused $Ga^+$ beam. (*c*) Boron nitride nanotube (BNNT) transmembrane device in a silicon nitride (SiN) membrane And a scanning electron microscopy image of the BNNT nanofluidic device. (*d*) A 2D silica nanochannel fabricated *via* the anodic bonding process. With high precision chemical etching, 2 nm-deep silica channels (atomic force microscopy image on the right side) with a width of 2 μm are fabricated under the application of a high voltage, 600 V. (*e*) An Å-slit device made by stacking 2D materials in the form of tri-layer stack containing bottom, spacer, and top layers. Panel *a* adapted with permission from Reference 34; copyright 2005, American Physical Society. Panel *b* adapted with permission from Reference 39; copyright 2018, American Chemical Society. Panel *c* adapted with permission from Reference 41; copyright 2013, Springer Nature. Panel *d* adapted with permission from Reference 44; copyright 2010, Springer Nature. Panel *e* adapted with permission from Reference 16; copyright 2018, Springer Nature.

## 2.1.3. 2D slit-like channels

Several top-down fabrication approaches are available to fabricate 2D slit-like channels including nanoimprint lithography, anodic bonding, and van der Waals assembly. For example, Xia et al. (43) fabricated silicon 2D nanoslits using nanoimprint lithography (NIL) and a melting reflow sealing technique. A pre-patterned array of 100 nm-wide silicon trenches created with NIL was exposed to an ultraviolet pulsed laser, which melted the thin surface layer of the silicon trenches. Molten silicon flows sideways and joins the neighboring pillars, forming the silicon channel, which could be further reduced to the dimensions of 20 nm in width and 60 nm in height after thermal oxidation (43). Later, Duan et al. (44) adopted and modified the etching followed by bonding method (45) in which previously the 2D channels made by anodic bonding of silicon and glass could stay stable only at a channel height of 20 nm due to the strong electrostatic forces between the substrates. Duan et al. (44) found that the electrostatic force can be reduced significantly by growing a 500-nm-thick silica layer on silicon, which permitted stable 2D channels with a height of 2 nm to be maintained (**Figure 2d**). In addition to silicon-based materials, remarkable progress has been achieved in nanofluidics using 2D-materials including graphene, hexagonal boron nitride (hBN), $MoS_2$, and other atomically flat materials. Recently, experimental methods to fabricate 2D slit-like channel *aka* 2D empty space which is only a few angstroms in height were developed by Radha, Geim and co-workers (46). Creation of 2D-empty spaces by selectively removing atomic layers is certainly a new direction of research for 2D materials (47). Here, the 2D channels are made by stacking 2D materials, similar to van der Waals heterostructures (46). As depicted in **Figure 2e**, they are tri-crystal structures where the bottom and the top layers define the walls of the 2D channel, while the spacer inbetween layers composed of parallel strips patterned by electron beam lithography (EBL), determines the width and height of the 2D channel (16).

## 2.2 Bottom-Up methods

In contrast to top-down methods, bottom-up methods involve making fluidic channels from intrinsic defects (e.g., vacancy pores), growing channel materials within devices (e.g., nanotubes), or creating channels by assembling the layers (laminates). Bottom-up approaches are more commonly used and usually result in multiple channels in a device. Several ways of making nano- or angstrofluidic systems using bottom-up fabrication protocols are described below (9).

### 2.2.1. 0D pore

Chemical vapor deposition (CVD) is the most common method used to fabricate continuous films of 2D materials. For instance, CVD graphene can make a 30-inch continuous film without any defects (48). Nevertheless, with proper control of the CVD growth conditions, it is possible to create intrinsic defects in the graphene layer. For example, Kidambi et al. (49) produced sub-2-nm 0D pores in an as-grown CVD graphene film *via* a lower than usual CVD growth temperature of 900 °C (**Figure 3a**). Another study by Griffin et al. (50) reported using a laser-assisted CVD process to generate a monolayer amorphous carbon (MAC) membrane with disordered non-hexagonal carbon rings from (**Figure 3a**), and sub-nanometer pore sizes, that permitted the flow of single lithium ions. Self-assembled aromatic monolayers have been demonstrated to form a monolayer carbon membrane with pores ~0.7 nm using electron irradiation (51).

### 2.2.2. 1D tubes

Excellent performance of biological protein channels (e.g. aquaporin), with both high permeability and selectivity, has inspired the construction of artificial membranes by incorporating aquaporin-like 1D structures (52). CNTs are promising systems that have been embedded into various silicon-based membranes to make fluidic chips. Holt et al. (53) reported a CNT-SiN hybrid membrane in which the vertically aligned CNTs were etched to open their entries as 1D channels with a channel diameter of 1.3 nm to 2 nm. However, the difficulty of controlling the individual CNT diameters and integrating them into the membrane without leaks can be challenging. Liu et al. (54) reported a microinjection technique to fabricate a CNT-liposome hybrid membrane, which has small pore size and well-defined pore geometry. In this method, a tip containing the solution of single-walled CNTs (0.8– 2 nm in diameter, all with the same chirality) is placed in proximity to the lipid layer. With a controlled injection rate of 1.0 μL/min, CNTs can effectively be inserted into the lipid bilayer. In other studies, Noy and coworkers (27, 55) produced CNT porins (CNTPs) in a liposome membrane. A monolayer of lipid molecules was coated on CNTs to induce their self-assembly into liposome walls, forming CNTPs with pore sizes ranging from 0.8 nm to 10 nm (27, 55), a typical image of which is shown in **Figure 3b** (27). Compared with the microinjection method, this self-assembly fabrication process is highly versatile for use in many applications (27).

**a** Intrinsic defects in CVD-grown monolayer carbon membrane

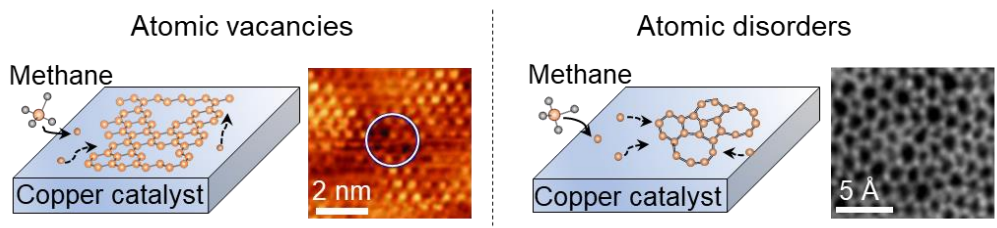

**b** Self-assembly of carbon nanotube porins into liposomes

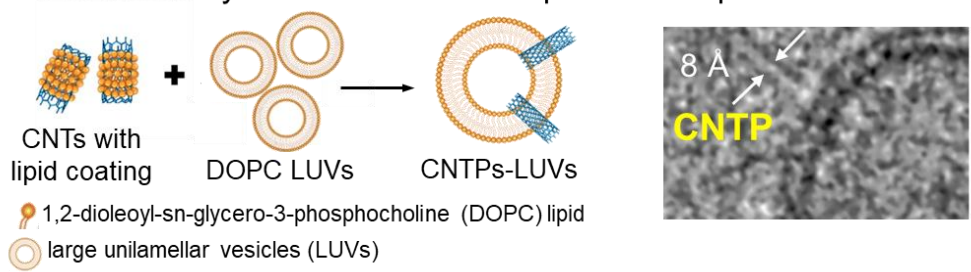

**c** Vacuum-assisted assembly of 2D MXene laminates

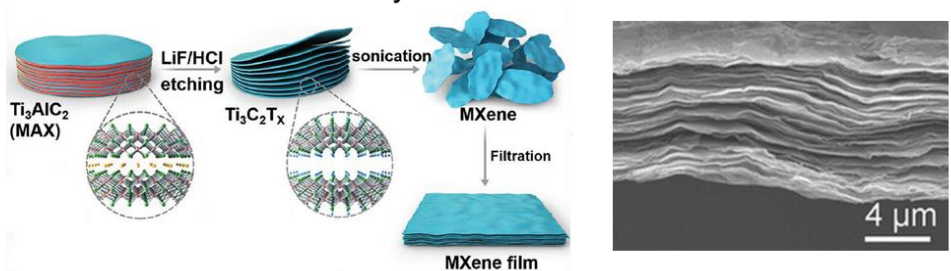

**Figure 3 State-of-the-art bottom-up fabrication methods for nano- and angstrofluidic systems.** (*a*, left): Monolayer graphene grown by a CVD process shows intrinsic nanoscale defects of carbon vacancies shown in white circle. (a, right) MAC membrane synthesized via a laser-assisted CVD method has intrinsic atomic-scale defects, such as five-, seven-, and eight-membered rings instead of regular hexagonal carbon rings. (b) Self-assembly of CNTPs into liposomes. On the right is a high magnification cryo-TEM image of the CNTP. (c) Vacuum filtration-assisted assembly of 2D MXene laminates. A cross-sectional SEM image of MXene laminate is shown on the right. Panel (*a, left*) adapted with permission from Reference 49; copyright 2018, John Wiley and Sons. Panel (*a, right*) Adapted with permission from Reference 50; copyright 2020, American Chemical Society. Panel *b* adapted with permission from Reference 27; copyright 2016, Springer Nature. Panel *c* adapted with permission from Reference 58; copyright 2017, John Wiley and Sons. Abbreviations: CNT, carbon nanotube; CNTP, carbon nanotube porin; CVD, chemical vapor deposition; DOPC, 1,2-dioleoyl-sn-glycero-3-phosphocholine; LUVs, large unilamellar vesicles; MAC, monolayer amorphous carbon; SEM, scanning electron microscopy; TEM, transmission electron microscopy.

### 2.2.3. 2D channels and laminates

Layer-by-layer assembly of 2D material sheets is a popular bottom-up approach for making 2D laminate channels. Pioneering research from Nair et al. (56) reported a graphene oxide laminate (interlayer spacing of ~1 nm) made with a vacuum-assisted filtration method; this work led to a surge in the development of 2D material-based membranes. Since then, viable fabrication methods such as evaporation-assisted assembly, pressure-assisted self-assembly, and spin casting have been used successfully to produce $MoS_2$, vermiculite, and MXene laminates (57). **Figure 3c** illustrates the synthesis process of 2D MXene membranes (58). In brief, the process involves selectively etching the middle aluminum layer of the $Ti_3AlC_2$ precursor by using liquid solutions of LiF and HCl which allows for easy delamination of $Ti_3C_2T_x$ layers under ultrasound sonication. The dispersed MXene nanosheets are stacked on a polymer support through vacuum filtration, resulting in a 2D lamellar structure (interlayer spacing of ~1.3 nm), which is shown clearly in the cross-sectional SEM image in **Figure 3c**. Readers are advised to consult other reviews (57, 59) to know more about the rapidly growing volume of work on 2D-laminates.

### 2.2.4. Open framework membranes

Open frameworks such as zeolites, metal organic frameworks, and covalent organic frameworks are often 3D entities with interconnected macropores and micropores, and are well-known for their applications as absorbents and catalysts (60). Recent reports have found that their molecular sieving and catalytic performance can be significantly improved by reducing their thickness to a few nanometers due to enhancement in molecular diffusion (61). Jeon et al. (62) demonstrated a seeded growth of 2D zeolitic nanosheets that have a final thickness of 4.7 nm (2.5 unit cells) with open pore entries of 0.6 nm. Compared with other methods of fabricating 2D zeolitic membrane which are usually multi-step, this direct synthesis method is cost-effective and gives a high yield with a micrometer lateral size. For further reading on the progress of this emerging field, we direct readers to other reviews (63, 64).

### 3. Toolbox of measurement techniques in the nanofluidics

In the previous section, we discuss the fabrication of subnanometer-scale fluidic channels. Such channels can show molecular transport deviating from the continuum, as well as unexpected phenomena that are yet to be understood. In order to observe and measure fluid behavior under sub-10-nm confinement, measurement techniques have been rigorously developed over the last decades, as described in this section.

### 3.1 Optical microscopy

Owing to its simplicity, optical microscopy has been one of the most prevalent methods to study fluid flows since the dawn of microfluidics (**Figure 4a**). However, the optical contrast is insufficient at the sub-10-nm scale to observe the liquid meniscus (liquid/gas interface) inside the nanochannels due to the short optical path along the confining dimensions. To overcome this limitation, optical microscopy has been adapted to suit nanofluidic measurements through using highly fluorescent dyes (44) or by adding a layer of SiN beneath the channels' surface to enhance the optical contrast between gas and liquid phases (65). Using optical microscopy, Duan and coworkers (66, 67) developed a method to measure the hydraulic resistance and slippage of single graphene nanochannels by fabricating hybrid nano-channels consisting of test and reference regions within a single channel of ~16 nm thickness (**Figure 4a**). Single-molecule localization microscopy is another notable development to overcome spatial and temporal resolution limits while using optical microscopy (68). Recent research demonstrated that single-proton transport can be visualized by a localized emission originating from the interaction between a proton and a boron vacancy in hBN (69).

Particle image velocimetry can visualize and measure flow velocity indirectly by optically observing flow-tracing particles dispersed in a liquid (70). By combining particle image velocimetry with a Landau−Squire nanojet, Bocquet and coworkers (71) devised a way to measure the pressure-driven flow rate through a single nanotube of 7 nm diameter. Using this approach, the permeability of CNTs and BNNTs with varying radii was extracted, establishing that the enhanced flows in narrow CNTs are due to unique radius-dependent hydrodynamic slippage.

## 3.2 Atomic force microscopy

Atomic force microscopy (AFM) is a well-known scanning probe technique capable of reaching atomic resolution. However, it is challenging to use conventional AFM to study liquid properties directly, due to the difficulty of approaching liquid surfaces without perturbing them. Because of the strong capillary forces between the liquid and the tip of the scanning probe, the liquid surface can be perturbed, which also affects the measurement accuracy. Nevertheless, the AFM technique was adapted to successfully visualize liquids by contouring them with graphene layers (72). Later on, the condensation and evaporation of water were imaged by using electrical polarization forces with lateral nanometric resolution, distinguishing a monolayer of water formed with apparent heights of ~2 Å (73). In another study, using scanning capacitance microscopy, a highly suppressed dielectric constant of water under sub-2-nm confinement was observed by Fumagalli et al. (20) using electrostatic properties (**Figure 4b**). Inside Å-slits, the out-of-plane dielectric constant of water varied from its bulk value of 80 to as low as 2 (with a confining height of ~7 Å), and the clear contrast between the dielectric constant of water confined at 1.4 and 10 nm was visualized (**Figure 4b**).

AFM was also used to quantitatively monitor the gas flow through vacancy defects in graphene membranes by measuring the change in volume of the graphene bulging (74). By improving the graphene's seal with the substrate using microcontainers of hBN walls, the limit of detection was brought down to being able to measure few tens of molecules crossing graphene membranes (75). In situ AFM was adapted for studying capillary condensation inside 2D Å-slits made from graphene and/or mica, validating the Kelvin equation at the atomic scale (76). Furthermore, combining AFM with electrochemical measurement techniques such as scanning ion conductance microscopy in a correlative manner offers characterization of not only liquids and gases but also ion transport (17, 77).

## 3.3 Transmission electron microscopy

Transmission electron microscopy (TEM) is a powerful tool to visualize events at the sub-nanometer scale (**Figure 4c**). Despite the advantage of high resolution, TEM requires the sample to be in a high vacuum, which is a hindrance to observing liquids due to evaporation. To facilitate the observation of fluids under TEM, researchers entrapped the liquid in confined structures either completely enclosed (78, 79), or with an open end (80). For instance, the confined water in a CNT with an inner diameter of 4 nm was visualized by TEM, leading to the conclusion that there was no clear meniscus distinguishing the liquid and gas phase (**Figure 4c**), which is in contrast to the conventional interface of water (78). Using TEM, ultrathin water films were observed to be highly stable under the high vacuum, which is attributed to high adhesion to the wall and reduced vapor pressure due to the confinement effect (80). Furthermore, in-situ TEM with liquid cells provides opportunities to study the dynamics of liquids in response to external stimuli including pressure, temperature, and electric or magnetic fields (81). The rapid improvements in imaging resolution and detector sensitivity for this technique can greatly benefit the in situ imaging of nano- and angstrofluidic channels to understand the dynamics of liquid/gas or liquid/solid interfaces, molecular ordering, electrical double layers, phase transitions, and even chemical reactions (11, 82).

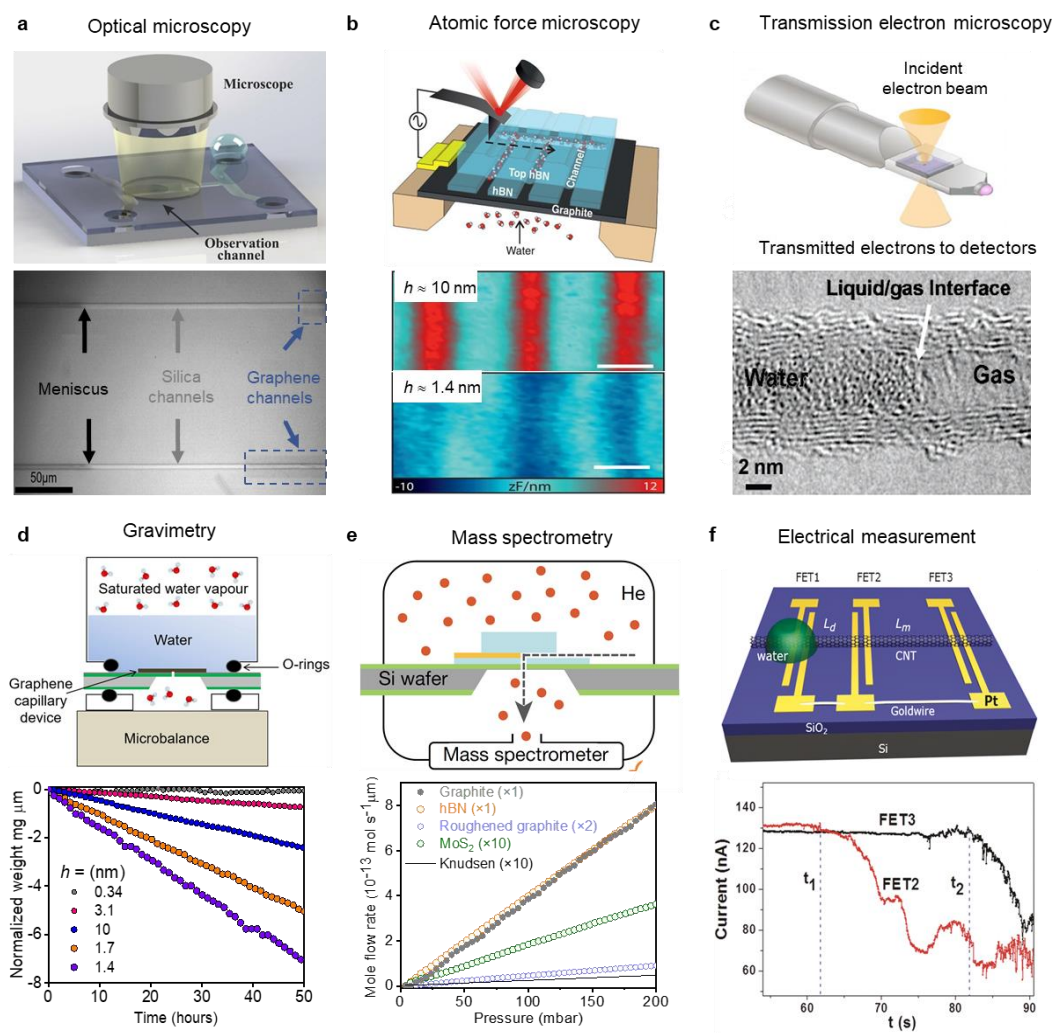

**Figure 4.** Toolbox of measurement techniques for nanofluidics. Schematics and experimental results for (*a*) optical microscopy of water flow in silica-graphene hybrid nanochannels (height 16 nm), (*b*) atomic force microscopy imaging of local capacitance of water inside nanochannels with confinement heights of 1.4 nm and 10 nm, (*c*) transmission electron microscopy visualizing a water/gas interface in a 4-nm CNT channel, (*d*) microgravimetry technique for measuring water flows through 2D slits with heights ranging from ~0.34 nm to ~10 nm, (*e*) mass spectrometry measuring gas (helium) permeation through 2D slits of height (~1.7 nm) with walls prepared from different crystals as labeled in the graph. The flow expected for Knudsen diffusion is shown by the solid black line. Plots in d and f were normailized by 1 μm, and (*f*) electrical measurement of water velocity inside CNTs using field effect transistor geometry. The velocity can be estimated from the distance between FET2 and FET3 and period from $t_1$ to $t_2$. Abbreviations: CNT, carbon nanotube; hBN, hexagonal boron nitride; FET, field-effect transistor. Panel *a* (top) adapted with permission from Reference 66; copyright 2016, Springer

### 3.4 Gravimetry

Gravimetry is a simple and robust quantitative measurement of an analyte's mass and was smartly modified to measure the liquid flows in nanochannels (56). A significant advance was made by Radha et al., (46) using microgravimetry with sensitivity of 1 μg to measure flows through highly uniform Å-slits. In a typical gravimetric measurement, the nanochannel device is mounted on top of a container partially filled with a liquid analyte placed on a microbalance, and the weight change is monitored (**Figure 4d**). The fluid flow through the nanochannel leads to a decrease in the weight of the analyte in the filled container, which is a direct measure of the flow. This is a highly effective method to observe the total amount of mass transport through the channels, but the obtained result is averaged value over entire membrane and cannot subtract the non-uniformity of the channels and pores. With microgravimetry, fast water transport with velocities up to 1 m/s were measured, which can be attributed to the high capillary pressure (up to 1,000 bar) and large slip lengths ($\delta$) on the graphene walls (**Figure 4d**). Additionally, this measurement allowed for the experimental establishment of the huge difference in $\delta$ between isoelectronic graphene ($\delta \approx 60$ nm) and hBN ($\delta \approx 1$ nm) sensitively (46). Such unconventional adaptation of mass transport techniques can certainly provide a new approach to measure anomalies in fluid dynamics.

### 3.5 Mass spectrometry

Mass spectrometry identifies analytes by sorting ionized gases in electric fields on the basis of their mass-to-charge ratio (85). It has been adapted to quantify gas permeation and selectivity through nanochannels, finding applications in gas separation, storage, and extraction processes. With membranes of high pore density, such as those made from 2D laminates (86), 1D CNT forests embedded across silicon nitride membranes (53), or high density 0D pores (87), it is feasible to achieve a sufficient signal-to-noise ratio. However, measuring permeation through single nanochannels or a few of them is challenging. The detection limit can be improved by performing measurements, including inputting gas to the membrane entirely in a vacuum. Keerthi et al. (16) demonstrated flows through single channels using a custom-made setup with two chambers, at the center of which the SiN$_x$/Si wafer with 2D Å-slits was placed. While one chamber was connected to the gas reservoir, the other was facing the mass spectrometer (**Figure 4e**). Their measurements showed that the speed of helium gas transport is orders of magnitude higher than that based on Knudsen theory due to negligible surface scattering resulting in frictionless gas flow through the channels made of the graphene and hBN (**Figure 4e**). Using the same setup, flows through 0D vacancy defects in WS$_2$ monolayers were measured to validate the applicability of the Knudsen equation down to 3 Å (88).

### 3.6 Raman spectroscopy

Raman spectroscopy is a non-invasive technique used to characterize nanomaterials. In nanofluidics, the low cross section of scattering from the tiny liquid volumes under confinement poses a challenge to obtaining a decent Raman signal. To overcome this challenge, the technique has been adapted to sense changes in the confining channel to measure the liquid flows (89). In particular, the radial breathing mode of CNTs can be monitored to understand the phase (gas/liquid/solid) of the filling fluid. For example, a large elevation of the freezing transition was observed inside a 1.05-nm CNT, enabling ice inside CNTs to be stable up to 138 °C (90).

## 3.7 Electrical measurements

Characterizing flows by electrical measurements is an interesting way to combine solid-state electronics with fluidic measurements. As an example, flow rate through a single nanotube was extracted by electrical measurements in field-effect transistor (FET) geometry. An individual CNT was demarcated into three FETs (**Figure 4f**), and the flow rate was estimated on the basis of the detected time for the electrical fluctuation (84). In this experiment, the flow rate of water through a CNT was found to decrease with increasing diameter, and a continuum breakdown was observed at a CNT diameter of 0.98 - 1.1 nm. Using the same principle as a solid-state transistor, fluidic transistors have been envisioned wherein one can influence the mobility of liquids and the analytes and ions dissolved in the liquids (91). Ionic flows can be an indirect probe for fluid flows because measuring electrical currents is easier than measuring fluid flow. The most common electrical measurements of ions are shown in **Figure 5b**, where the ionic properties are extracted from the current-voltage (I-V) characteristics. Figure 5 provides an explanation of the corresponding measurement technique. In what follows, we describe various driving forces inducing ionic movement (**Figure 5**).

## 4. Ionic Transport

Understanding ion transport in nano and angstrom scale channels has practical relevance in applications such as membrane desalination, blue energy, supercapacitors and batteries, as well as in understanding ionic flow through biological channels (14, 29, 31, 32). Compared with water and other fluids, ions have long-range electrostatic interactions. The Coulomb interactions between ions and the nano/angstrom-channel walls can follow a complex physics where several factors are important including the dimensions of the channels, and the composition and properties of the channel walls, the concentration and pH of the electrolyte (8). It is essential to understand several characteristic ionic interaction lengths (see the sidebar titled Ionic interaction lengths) at play in the electrokinetic phenomena of nanofluidic systems, such as Debye length ($\lambda_D$), Bjerrum length ($l_B$), Gouy-Chapman length ($l_{GC}$) and Dukhin length ($l_{Du}$), which are summarised in the **Supplemental Material**. As we approach the nano- and angstrom-scale, the

**Ionic interaction lengths**

**Debye length:**

$$\lambda_D = \frac{1}{\sqrt{8\pi l_B C}}$$

**Bjerrum length:**

$$l_B = \frac{z^2 e^2}{4\pi \varepsilon_r \varepsilon_0 k_B T}$$

**Gouy-Chapman length:**

$$l_{GC} = \frac{1}{2\pi \Sigma l_B}$$

**Dukhin length:**

$$l_{Du} = \frac{|\Sigma|}{C}$$

Where, $C$ is concentration of the salt in the reservoirs, $\varepsilon_r$ is the dielectric constant of the medium, $\varepsilon_o$ is the permittivity of free space, $k_B$ is the Boltzmann constant, $T$ is the temperature, $\Sigma$ is the channel's surface charge, $e$ is the elementary charge, $z$ is the ion valency.

channels' confining dimension becomes similar to these ionic lengths and ionic dimensions (**Table 1** in the **Supplemental Material**). For a comprehensive overview of ion transport basics, please see the reviews (8, 92, 93).

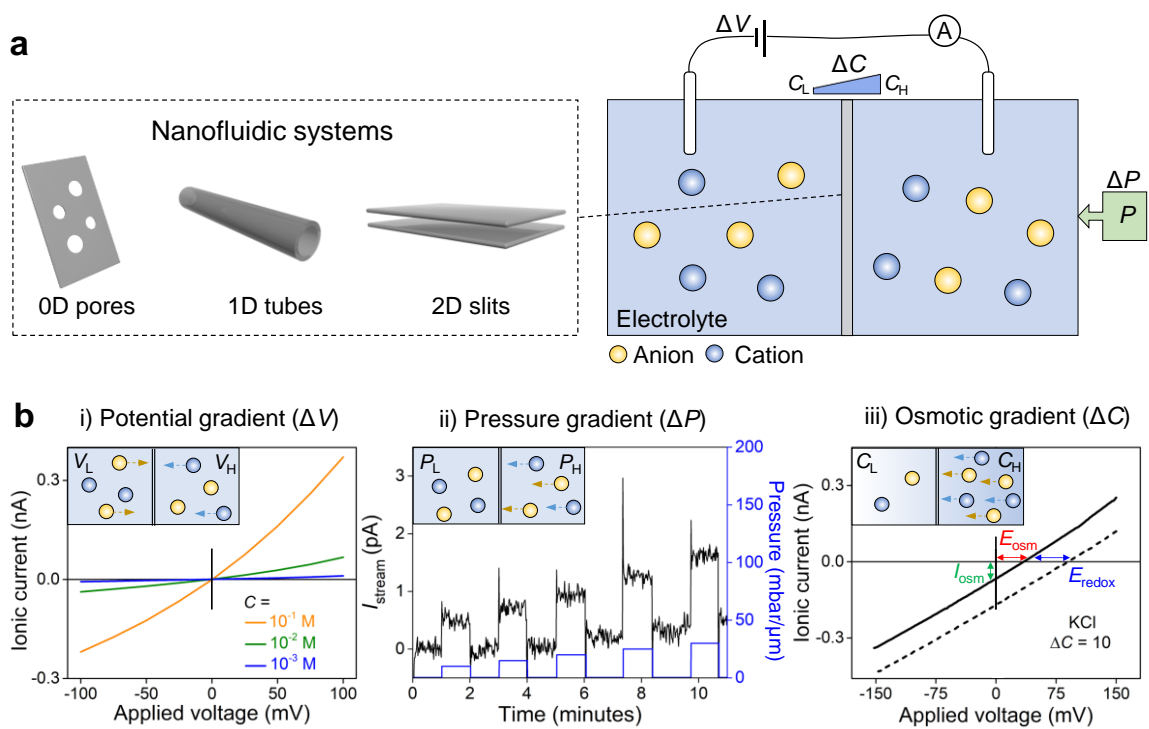

**Figure 5. Stimuli-driven ion transport in nanofluidic systems.** (*a*) Schematic illustration of the experimental measurement set-up of ion transport through different nano- and angstrofluidic systems including 0D pores, 1D tubes, and 2D slits. (*b*) Representative ion measurement curves obtained under different stimuli ($\Delta V$, $\Delta P$, $\Delta C$). i) Current-voltage (*I-V*) characteristics of a 2D Å-slit ($h = 6.8$ Å) for various KCl concentrations. In the inset, cations (blue spheres) migrate from the high potential ($V_H$) towards the low potential ($V_L$) compartment while anions (yellow spheres) transport against $\Delta V$, resulting in the generation of electrical current. ii) Streaming current variation (black curve, left y-axis) with increasing $\Delta P$ (blue curve, right y-axis) inside a 2D-slit device filled with 0.3 M KCl solution. Inset, cations and anions flow from the high pressure ($P_H$) to the low pressure ($P_L$) compartment. iii) For the same 2D-slits, *I-V* curves of KCl salt under 10-fold $\Delta C$. The black dashed *I-V* line intersection with the x-axis shows total cell potential, whereas osmotic potential ($E_{osm}$, solid black line) could be deduced from the total cell potential by subtracting the electrodes' redox potential ($E_{redox}$). The osmotic current ($I_{osm}$) is the intersection of the solid black line with the y-axis. In the inset, both cations and anions diffuse from the high concentration ($C_H$) to the low concentration ($C_L$) compartment. Panel (*b i*) adapted with permission from Reference 95; copyright 2017, American Association for the Advancement of Science. Panel (*b ii*) adapted with permission from Reference 96; copyright 2021, Springer Nature.

Charged nanochannel walls attract ions of opposite charge and repel ions of same charge to conserve the electroneutrality of the solid-liquid interface, thus forming an electrical double layer (EDL). This layer comprises an immobile Stern layer and a mobile diffuse layer. The electric potential decreases with the distance from the wall until the electro-neutrality of the solution is established at the end of the diffuse layer. As a rule of thumb, the EDL thickness is usually four times the $\lambda_D$, but some studies consider EDL thickness to be the same as $\lambda_D$ (94). When the EDL becomes comparable in thickness to the confining dimension of the channel, the ion movement inside the channel will be dependent on the size and composition of the EDL inducing charge selectivity (92). This effect is seen in nanofiltration and reverse osmosis membranes where small pores display higher selectivity and separation of dissolved ions (97). Ion selectivity is frequently correlated with the overlapping of the EDLs formed at the opposite walls in nanochannels. However, nanopores with dimensions much larger than the EDL also are reported to show good selectivity. For instance, using large pores (diameter, 10-100 nm) without EDL overlap, Poggioli et al. (98) showed that the principal factor controlling the selectivity is the $l_{Du}$ rather than the EDL overlap or $\lambda_D$. It is worth noting that $l_{Du}$ has much higher values than $\lambda_D$ at typical salt and surface charges, and hence larger pores can be designed for selectivity.

The EDL overlap in the nanoscale affects electrokinetic phenomena, such as electroosmosis, electrophoresis and streaming (see the **Supplemental Material**). For example, in capillary electrophoresis, it is well-known that the electro-osmotic flow (EOF) greatly affects ionic and molecular movements at the microscale. However, at the nanoscale, where the EDL overlaps, the effect of electroosmosis changes drastically (44). In the case of super-hydrophobic nanochannels, where water does not have affinity to the walls, the EOF increases by several orders of magnitude (99-101). Whereas, in the case of hydrophilic channels the EOF becomes negligible and transport is primarily through electromigration (102, 103). Moreover, EDL overlapping with or without charge asymmetry results in non-linear electrical responses such as current rectification or Coulomb blockade (38, 104).

In the following sections, we discuss ion movement (**Figure 5**) through various nanofluidic systems under different stimuli (electrical, pressure, osmotic) and emphasize the applications sought from such systems in relation to energy harvesting, nanofiltration and smart biomaterials. Although not an exhaustive summary of the published results, we draw overall conclusions to help future advances in this domain.

## 4.1 Ion flow driven by electrical stimulus

---

**Electrokinetic Mobilities**

**Apparent mobility:**

$$\mu_{app} = \mu_{eof} + \mu$$

**Electrophoretic mobility:**

$$\mu = \frac{q}{6\pi\nu r_s}$$

**Electro-osmotic mobility:**

$$\mu_{eof} = \frac{1}{2hE}\int_0^{2h}\frac{\varepsilon_r\varepsilon_0\zeta E}{\nu}\left(1 - \frac{\psi(y)}{\zeta}\right)$$

**Counter-ion mobility in surface-charge-governed regime:**

$$\mu_s = \frac{Gl}{2w\Sigma}$$

In these equations, $q$ is the charge of the solute (or ion), $\nu$ is the viscosity of the medium, $r_s$ is the radius of the solute, $E$ is the electric field, h is the height of the channel, l is the length of the channel, w is the width of the channel, $\varepsilon_r$ is the dielectric constant of the medium, $\varepsilon_0$ is the permittivity of free space, G is conductance, $\Sigma$ is the channel surface charge $\zeta$ is the zeta-potential, and $\psi(y)$ is the equilibrium potential at position $y$ perpendicular to the wall.

An electric potential applied across nanochannels leads to the flow of ions across those nano channels. The speed of the ions, expressed by their mobility (see sidebar titled Electrokinetic Mobilities), and the selectivity between the different cations and anions are crucial for applications in logic devices as well as supercapacitors. Trends of different reported enhanced or suppressed K$^+$ mobility with respect to the bulk systems ($\mu_{confined}/\mu_{bulk}$) are shown in **Figure 6a**. Different literature results are presented (**Figure 6**) which are classified on the basis of the dimensionality and the material of the nanofluidic system as well as whether the results are experimental or simulated.

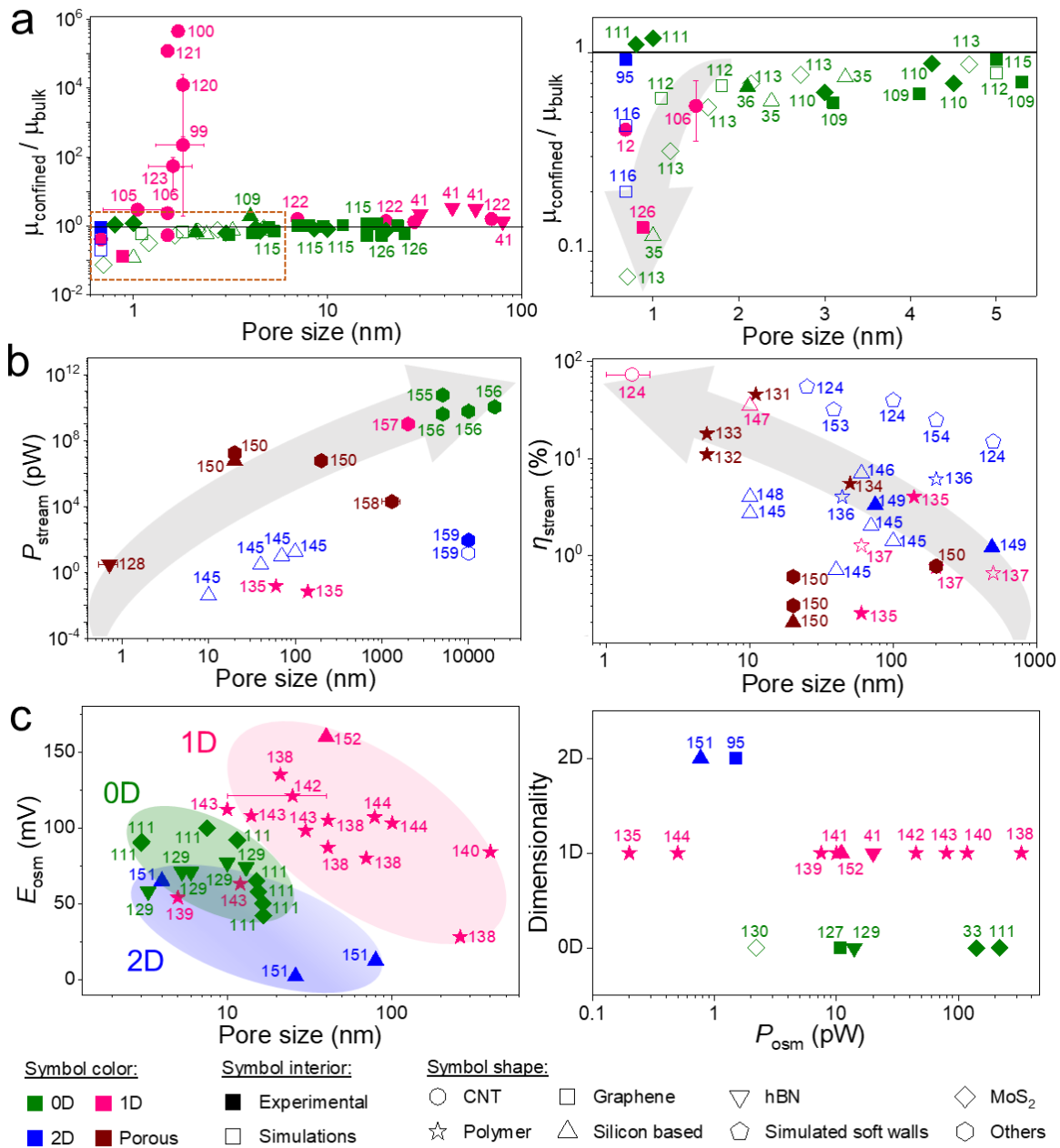

**Figure 6:** Ionic transport in nanofluidic systems under electrical, pressure and osmotic stimuli. (*a*, left) Ratio of confined to bulk mobilities for K$^+$ ions plotted as a function of the pore size. (a, right) A zoomed view for pore sizes of sub-5-nm (the area indicated by a red dashed box in the left graph). As the channel size increases, the confined K$^+$ mobility approaches its bulk value. For some literature studies shown on the graph, where K$^+$ mobility was not directly given and there was no selectivity between ions, we estimated the enhancement in K$^+$ mobility from the conductance enhancement of KCl. (*b*, left) The streaming power ($P_{stream}$) increases as a function of the pore size. (b, right) The streaming efficiency ($\eta_{stream}$) of the system decreases with pore size. (c, left) Variation of the osmotic potential ($E_{osm}$) as a function of the pore size when a 1000-fold salt concentration gradient ($\Delta C$) was used. (c, right) Variation of the osmotic power ($P_{osm}$) when $\Delta C = 10\text{-}1000$. The different properties of the chosen nanofluidic systems such as its dimensionality (0D, 1D, 2D, porous) and material composition (CNT (12, 99, 100, 105, 106, 120-124), graphene (95, 109, 112, 115, 116, 125-127), hBN (41, 128, 129), MoS₂ (33, 110, 111, 113, 130), polymers (40, 131-144), silicon based (35, 36, 109, 145-152), simulated soft walls (124, 153, 154) and others including metal (155, 156), glass (157-159), alumina (150) and amino silanes (150)) as well as the type of the study (experimental or simulations) are represented by the symbol colour, shape and interior filling, respectively. The horizontal error bars represent the range of the pore sizes given in the specific literature report, while the vertical error bars indicate the variation in the ionic mobility and conductance reported by the corresponding studies. Data in panels (a) and (c) are from single nanofluidic systems; porous membranes were excluded for the reasons specified in the text. In panel b, porous membranes were taken into consideration as there are only a limited number of studies with single channel systems. The grey arrows in the background are a visual guide of the variation of the measurable quantity ($\mu_{confined}/\mu_{bulk}$, $P_{stream}$, $\eta_{stream}$, and $E_{osm}$) with the variation of the pore size. The green, pink and blue shaded areas in panel (c, left) correspond to 0D, 1D and 2D nanofluidic systems, respectively. Abbreviations: CNT, carbon nanotube; hBN, hexagonal boron nitride.

1D nanofluidic systems have been studied extensively. The general trend has been the enhancement of ion (K⁺) mobility for sub-5-nm 1D systems, and as the channel size increases, the confined mobility values approach bulk values (**Figure 6a**). This increase in the mobility was attributed to higher slippage and the variation of the electrokinetic phenomena at such small dimensions. Specifically, as the dimension of the nanofluidic system reduces to Angstrom-scale, the EOF increases resulting in an increase of the apparent mobility (105). For example, Pang et al. (100) showed that the mobility of ions flowing through CNTs of 1.7 nm diameter, increased by more than four orders of magnitude relative to the bulk. The increase in the speed of ions is not due to an increase in the electrophoretic mobility, but rather an increase in the EOF inside the CNTs, which carried the ions with water flux. Coupling EOF with electrophoretic ion flow was also reported by Lee et al. (99), and Yao et al. (101), where Yao et al. (101) observed that the apparent mobility of K⁺ ions was enhanced (2.4 times) with respect to their bulk mobility despite a decrease (40%) in their electrophoretic mobility. As shown in **Figure 6a**, the mobility in a few 1D nanosystems showed suppression rather than enhancement (see sidebars Osmotic potential and titled Theoretical Conductance and Mobility Calculation). For example, using ultra small CNTs (diameter = 8 Å), Tunuguntla et al. (12) observed that an increase in water flow inside a CNT did not result in an increase in salt conductivity. This might be attributed to the charged entrance and to the possible confinement of the ions. Similarly, Geng et

## Osmotic Potential

To obtain the osmotic potential ($E_{osm}$) experimentally, we need to subtract the redox potential ($E_{redox}$) when standard Ag/AgCl electrodes are used, instead of reference electrodes. The redox potential arises due to the difference of chloride concentration between the two cells from the overall observed open circuit potential ($E_{oc}$)

$$E_{osm} = E_{oc} - E_{redox}$$

$$E_{redox} = \frac{RT}{zF} ln(\frac{\gamma_H C_H}{\gamma_L C_L})$$

where, $R$ is the universal gas constant, $F$ is Faraday's constant, T is the temperature , z is the valency of the ion and $\gamma_{H/L}$ and $C_{H/L}$ are the activity coefficient and the concentration of chloride anion on the high (H) and low (L) concentration sides.

al. (106) reported a decrease in salt conductivity inside CNTs. They explained this phenomenon by the existence of low-conductance sub-states for charged CNTs and on/off transitions for uncharged CNTs, which is similar to what happens in biological pores. In addition to the mobility, the selectivity of an ion is another important parameter. In CNTs, ionic selectivity of up to more than 99% has been observed for both cations and anions (12, 107). The reason behind the high selectivity was interpreted to be the Donnan or charge exclusion at the entrance of the CNT and this can be tested by measuring the selectivity in two different pH solutions of the same salt concentration, as pH will affect the ionization of the functional molecules at the entrance. Increases in the concentration of the electrolyte and the size of the pores can both decrease the selectivity (108). Apart from CNTs, ionic conductance enhancement was also seen for other 1D systems. For instance, Siria et al.(41) reported four times enhancement of KCl conductance in hBN nanotubes (**Figure 6a**).

In contrast to 1D tubes, 0D pores and the 2D slits mostly showed suppression of $K^+$ mobility, with few exceptions (**Figure 6a**) (36, 109-115). The mobility suppression is evident in the move from nano- to angstrom-scale where the effect of $l_{Du}$ and $\lambda_D$ becomes much more pronounced. 2D slits (height of ~ 6.8 Å) made from graphene, hBN, or $MoS_2$ showed a suppression in the mobility of an anion ($Cl^-$) as well as cations ($Ca^{2+}$, $Mg^{2+}$, $Fe^{3+}$, $In^{3+}$, $Al^{3+}$) compared with expected bulk mobility with the exception of $Li^+$ and $Na^+$ ions which showed slight (20- 40%) enhancement (95). The suppression of the mobility of the cations was proportional to the hydrated diameter ($D_H$), which is attributed to the squeezing and partial dehydration of the large ions as they pass through the 2D slits. Later, Yu et al. (116) performed molecular dynamics (MD) simulations on Å-slits with similar dimensions and confirmed the distortion of the hydration shell around ions when entering the slit, leading to mobility suppression. Remarkably, similar findings have also been reported in several other MD simulations (**Figure 6a**) (35, 36, 112, 113, 117-119).

## Theoretical Conductance and Mobility Calculation

**0D pore:** $\qquad G = \sigma \left[ \frac{4l}{\pi d^2} + \frac{1}{d} \right]^{-1}$

For ultrathin membranes where channel resistance (first term of above equation) is much smaller than the access resistance (second term of above equation):

$$G = \sigma d$$

Note that for the 0D pores fabricated from ultrathin 2D materials presented in figure 6, the theoretical bulk conductance was calculated using the simplified formula

**1D tube:** $\qquad G = \sigma \left[ \frac{\pi d^2}{4l} \right]$

**2D slit with 'n' channels:**

$$G = \sigma \left[ \frac{whn}{l} \right]$$

The electrophoretic mobility calculation from conductivity and from osmotic potential is:

$$\sigma = F(z^+ \mu^+ C^+ + z^- \mu^- C^-)$$

$$\frac{\mu^+}{\mu^-} = -\frac{z^+}{z^-} \frac{ln\left(\frac{C_H}{C_L}\right) - z^+ FE_{osm}/RT}{ln\left(\frac{C_H}{C_L}\right) - z^- FE_{osm}/RT}$$

In these equations, $G$ is conductance, $\sigma$ is bulk ionic conductivity, $d$ is pore diameter, $\mu$ is the electrophoretic mobility, $l$, $w$, $h$ and $n$ are the length, width, height and number of channels, respectively, F is Faraday's constant, $E_{osm}$ is the osmotic potential, $C^+$ & $C^-$ are concentration of anion and cation respectively, $C_H$ and $C_L$ are the concentration of the salt on the high ($H$) and low ($L$) concentration sides, $z^+$ and $z^-$ are the cation and anion valencies, respectively. T is the temperature, and R is the universal gas constant.

The ions' mobility suppression could be due to several reasons, such as easier formation of ion pairs inside confined channels, the electrostatic interaction of the ions with the charges on the channel walls (known as surface trapping), or the balance between entropy and hydration energy (160). Careful theoretical simulations have demonstrated the difference in ion mobility as a factor in the ion position (near or away from the wall) inside the channels (160). For sub-5-nm pores, low ionic mobility was predicted due to ion partitioning arising from steric hindrance and the layered liquid structure in the pore axial direction. MD simulations on sub-2-nm $MoS_2$ nanopores by Pérez et al. (113) showed a correlation between the mobility suppression and the decrease in the pore diameter. Similar mobility suppression with decreased channel size has been reported experimentally by Ma et al. (36) for silicon nanopores (2-8 nm). On the contrary, Jung et al. (161) reported enhancement of NaCl conductance in multilayer graphene 2D-nanoslits (heights of 3.6, 10, and 50 nm) up to two orders of magnitude more than in silica 2D-nanoslits. Indeed, in both silica and graphene channels they observed enhancement of NaCl conductivity with respect to bulk, which was attributed, as in the case of CNTs, to the coupling with the EOF and to the high slip length of hydrophobic graphene.

## 4.2 Ion flow driven by pressure stimulus

Applying a pressure difference $\Delta P$ across the nanofluidic system to produce energy is called electrokinetic energy conversion (EKEC). Although the principle of EKEC was first described in the 1960s by Morrison & Osterle (162), in the last two decades, EKEC has gained further attention as a viable energy generation method for micro- and nanofluidics. Mechanical energy is converted into electric energy when the applied pressure forces the water molecules to flow through the nanochannels dragging the ions in the diffuse EDL layer with them. The passage of ions with different mobilities across the nanofluidic channels generates a streaming current, $I_{stream}$. High selectivity between the oppositely charged ions will generate a high $I_{stream}$ for the same applied $\Delta P$. Alternatively, when the current is supplied to an external load, a potential difference (streaming potential, $E_{stream}$) is generated across the channel, giving rise to an electrical streaming power ($P_{stream}$). The main parameters controlling the $E_{stream}$ are the surface charge, the electrolyte and its concentration, and the channel dimensions (see the formula in the sidebar titled Pressure Parameters). Furthermore, the larger the $D_H$ of the ion is, the higher the $E_{stream}$ ($Li^+ > Na^+ > K^+$) (163). Both $E_{stream}$ and $I_{stream}$ increase linearly with the applied $\Delta P$. Using low concentrations of the salt enhances the selectivity benefiting from high $\lambda_D$ and EDL overlap, and hence salt concentration is a critical parameter in EKEC (150). The main limitation of this energy conversion method is its low output power, and most research is still focused on simple nanofluidic devices for better understanding how to further develop the method.

The first attempts to generate electrical power by using pressurized water or salt were made on hard microporous materials such as glass and metal (156, 158). Pressuring water inside large metal orifices led to dragging the protons in the EDL, generating $P_{stream}$, though the efficiencies were low (**Figure 6b**). The main challenge was the fabrication of several parallel channels that can withstand high pressure to produce reasonable $P_{stream}$ and streaming efficiency ($\eta_{stream}$) (157). Decreasing the salt concentration and increasing the pore size resulted in higher $P_{stream}$ (145). However, pores that are too large will generate turbulent flow and inconsistent EDL and narrow pores will result in high entry resistance, both leading to decreased $P_{stream}$ and $\eta_{stream}$. From the literature reports, the $P_{stream}$ generated by EKEC increased with larger pore size $\eta_{stream}$ showed the opposite trend (**Figure 6b**). Moving from micro- to nano-channels is thus beneficial for higher efficiencies due to the increase in EDL overlap and the selectivity offered by the nanofluidics, which results in less energy dissipation. Moreover, increasing the slip length can increase the efficiency (124, 153), for example, CNTs with their hydrophobic and atomically smooth walls have large slip length and can show high $\eta_{stream}$ due to the decrease in the fluidic impedance and the resulting increase in the $I_{stream}$ (124). Another way to increase the $\eta_{stream}$ is by increasing the charge inside the channels. It must be noted that increasing the surface charge can enhance the permselectivity of a nanochannel, thus increasing $E_{stream}$ and $\eta_{stream}$, but it also leads to a reduction in the slippage due to increased affinity of water molecules with the walls of the nanochannels. Consequently, a moderate surface charge in an optimal range results in the highest possible $\eta_{stream}$ (164). This was confirmed by the results of Haldrup et al. (131), who obtained approximately three times higher efficiency by using membranes made from nitrocellulose and sulfonated polystyrene with ~3.4 times less surface charge than by using Nafion membranes (133, 134).

Let us point out that there are large number of theoretical studies as compared with experimental studies **Figure 6b**, the theoretical studies estimated better efficiencies than what was practically achieved. This disparity could be due in part to the materials used in the experiments. However, the overall efficiencies are mostly less than 10%, and this discrepancy between theoretical and experimental studies is elaborated in the cited references.

Coupling the electrical and potential stimuli is the subject of several studies with a target to design smart channels with possible gating effects (see Section 4.4). For example, Mouterde et al. (165), demonstrated that pressure-driven ion transport could be successfully gated by applying an electric field across the Å-slits of graphene and hBN, revealing electrohydrodynamic transistor-like behavior. Application of a small voltage resulted in an increase of the $I_{stream}$ by up to 20 times. MD simulations explained the origin of such non-linear coupling between potential and pressure to be the reduction of the total potential energy barrier due to the density variations of ions and water molecules in the channel (166). Recently, Marcotte et al. (42) demonstrated CNTs ($R \approx 2.0 \pm 0.6\ nm$, $l \approx 1\ \mu m$) to be mechanoreceptors controlled by small voltage bias. Contrary to the 2D slits which showed linear dependence of the ionic current on the applied pressure, the 1D tubes showed a quadratic pressure dependence. This variation was attributed to the ultra-fast water movement inside the CNT due to large slippage. Future mechanosensitive devices made from different materials are very promising for the development of smart gated nanofluidic

**Pressure Parameters**

**The streaming potential is:**

$$E_{stream} = \frac{\varepsilon_r \varepsilon_0}{\sigma \nu} \Delta P \zeta$$

**The streaming current is:**

$$I_{stream} = -\varepsilon_r \varepsilon_0 \zeta w h \Delta P / \nu l$$

**The streaming efficiency is:**

$$\eta_{stream}$$
$$= \frac{P_{output\ (electrical)}}{P_{Input\ (hydrodynamic)}}$$
$$= \frac{2\pi(\varepsilon_r \varepsilon_0 \zeta)^2}{\Sigma \nu A}$$

In these equations, $\varepsilon_r$ is the dielectric constant of the medium, $\varepsilon_0$ is the permittivity of free space, $\sigma$ is bulk ionic conductivity, $\nu$ is the viscosity of the medium, $\zeta$ is the zeta potential, $h$ is the height of the channel, $w$ is the width of the channel, $l$ is the length of the channel, $\Sigma$ is the surface charge. $\Delta P$ is the pressure difference applied on the nanofluidic system, and $A$ is the area of system.

systems that could be used for ion gating and molecular delivery, as in the case of biological channels.

### 4.3 Ion flow driven by osmotic stimulus

Ion diffusion with a concentration gradient ($\Delta C$) transforms the free Gibbs energy into electrical energy or osmotic power, and this process is known as reverse electrodialysis. The higher the $\Delta C$ and the selectivity, the higher the osmotic potential ($E_{osm}$) and the obtained osmotic power ($P_{osm}$) (see the sidebar titled Osmotic Parameters). Owing to the simplicity of the process, osmotic power generation is more popular than EKEC. For instance, mixing sea water and river water can generate electric power (41, 140), hence the name "blue energy". Ion exchange membranes have been widely used despite their drawbacks, including low $P_{osm}$ due to high membrane resistance, fouling, and inadequate mass transport. These problems arise from their high surface charge combined with their large membrane thickness and their channel openings being similar to ionic diameters (less than 1 nm) (31). Studying this phenomenon using single nanofluidic systems, rather than membranes, permits better understanding of the mechanism due to precise control of the dimensions and the charge of the system, paving the path to future scaling for high osmotic power delivery.

Single-channel nanofluidic systems show much higher osmotic power density ($P_{density}=P_{osm}$/opening area) than multi-channel membrane systems ($10^3$–$10^6$ W/m$^2$ vs 5 W/m$^2$) (31). This could be due to better optimization of the single system, but we also cannot rule out the artefact introduced in comparison by the area term (167). In practical applications, for membranes with several parallel conduits, a minimum distance is required between the channels, and when several pores are nearby the entry resistance effect needs to be taken into account (168). Hence, in the case of single-channel systems, dividing the $P_{osm}$ with the area of the channel opening is an overestimation of the osmotic power density. Indeed, Su et al. (169) observed that the power density increases non-linearly with the pore number and even declines after certain threshold, as the number of nanochannels are scaled up. They explained this by a decrease in the selectivity due to the crowding of ions which induces strong concentration polarization. Therefore, we focus on comparing the absolute $P_{osm}$ per single nanochannel (**Figure 6c**).

$P_{osm}$ depends on both $E_{osm}$ and osmotic current ($I_{osm}$). The $E_{osm}$ can be maximized by increasing the $\Delta C$ and ion selectivity (see the formula in the sidebar titled Pressure Parameters). **Figure 6c** (left) shows $E_{osm}$ values reported in the literature for a given $\Delta C$ of 1,000. Overall, $E_{osm}$ decreased with increasing channel opening, consistent with a decrease in selectivity. This highlights the importance of nanofluidic systems for harvesting higher osmotic energy. We categorize several literature reports according to the nanofluidic conduit's composition and dimensionality. For similar pore sizes, the 1D systems, mainly track-etched pores made from different polymers (138, 142, 143), show higher $E_{osm}$ than 0D $MoS_2$ and hBN nanopores (111, 129). The 2D nanoslits (made from silicon) (151) showed the lowest $E_{osm}$ (**Figure 6c**). However, let us note that the other component of the $P_{osm}$, that is, $I_{osm}$ is also sensitive to the dimensions of the channel (area and length) and the surface charge (see formula in the sidebar titled Osmotic Parameters). These would change the above discussed trends of dimensionality, for example, 2D slits with high surface charge showing much higher $P_{osm}$ than other systems (170).

To obtain a higher output of $P_{osm}$, the length of the nanochannel is an important parameter. The channel should be long enough to permit sufficient selectivity and short enough to avoid unfavorably large resistances (108, 152). Several reports discuss the effect of the channel length or membrane thickness on the generated power. For example, $P_{osm}$ of alumina nanochannels increased as the channel length decreased until an optimal length, below which the $P_{osm}$ decreased as the short channels showed ion concentration polarization governed regime. (171). Our summary of values found in the literature in **Figure 6c** (right) shows that 0D $MoS_2$ nanopores (111, 172), which have shorter lengths than 1D and 2D-nanochannels, have a higher maximum $P_{osm}$. All of these reports confirm the complex mechanisms controlling the $P_{osm}$. It is most likely that shorter channels with high selectivity or those systems with enormously high surface charge, yet showing no hindrance to ion movements, would lead to better output. However, the main challenges that need to be overcome for future large-scale applications of nanofluidic systems are mechanical robustness and scalability.

**4.4 Smart Ionic Devices:**

Development of smart devices that allow for modifying the ion/molecular transport inside nanochannels with external stimuli has been widely explored in the last few years (173). Through specific functionalization, the nanochannels can be made responsive to stimuli such as light, pH, temperature, voltage or the presence of certain ions and molecules. The main mechanisms for tuning the ionic selectivity are modifying the charge distribution, changing the wettability, and inducing conformational changes on the walls of the channels (**Table 2** in the **Supplemental Material**).

## Osmotic Parameters

**The Reversal/osmotic potential is:**

$$E_{osm} = (t^+ - t^-) \frac{RT}{F} \ln\left(\frac{\gamma_{C_H} \cdot C_H}{\gamma_{C_L} \cdot C_L}\right)$$

The cation and anion transference numbers are:

$$t^+ = \frac{z^+ C^+ \mu^+}{z^+ C^+ \mu^+ + z^- C^- \mu^-}$$

$$t^- = \frac{z^- C^- \mu^-}{z^+ C^+ \mu^+ + z^- C^- \mu^-}$$

**The osmotic current is:**

$$I_{osm} \approx \frac{2\pi r \Sigma}{l} \frac{k_B T}{\nu l_B} \log(\frac{C_H}{C_L})$$

**The maximum osmotic power is:**

$$P_{osm} = \frac{1}{4} E_{osm} I_{osm}$$

In these equations, r is the radius of the opening of the nanotube, F is Faraday's constant, T is the temperature, $\gamma_{H/L}$ and $C_{H/L}$ are the activity coefficient and the concentration of the salt on the high (H) and low (L) concentration sides. $z^+$ and $z^-$ are the charges of the cation and the anion, $C^+$ and $C^-$ are the concentrations of the cation and the anion, $\mu^+$ and $\mu^-$ are the mobilities of the cation and the anion, respectively, $\Sigma$ is the channel's surface charge, $l$ is the length of the channel, $\nu$ is the viscosity of the medium, $k_B$ is the Boltzmann constant, $l_B$ is the Bjerrum length, $E_{osm}$ and $I_{osm}$ are the osmotic potential and osmotic current, respectively.

Several applications are demonstrated with smart ionic gating, such as i) active ionic transport (against the $\Delta C$), ii) tuning the downhill ionic flow (along the $\Delta C$), that is, ionic flow gating and iii) modulation of ionic rectification, where the direction of ionic flow can be defined and thus the device acts as an ionic diode. For a summary of the latest stimuli, receptor mechanisms, and applications of the responsive nanochannel systems, see **Table 2** in the **Supplemental Material**.

Light is the most popular stimulus for ionic control in nanochannels owing to its ease of implementation as well as the desire to mimic light-gated protein channels, which are central to signal transmission in nature (e.g., channel rhodopsins). Shining light has been shown to generate higher osmotic energy, gate the ionic flows and change the ion selectivity (**Table 2** in the **Supplemental Material**). An interesting application of light as stimulus is uphill (active) transport mimicking biological channels (e.g.,: $Na^+/K^+$ ATPase) (40). The active transport is usually confirmed by the inversion of the sign of the zero-volt current. Active transport under higher $\Delta C$ is still challenging. Xiao et al. (174) were able to show active transport against larger $\Delta C$ up to 5000 folds by illuminating a membrane comprising a carbon nitride nanotube (diameter ≈ 30 nm, length ≈ 500 nm) with white light (**Table 2** in the **Supplemental Material**). The light induces a surface charge redistribution leading to the separation of electron hole pairs in the membrane and resulting in ion pumping.

Other types of stimuli such as thermal, electrical and pH are also explored in the literature for gating ionic flows (**Table 2** in the **Supplemental Material**). The pH stimulus showed very high ion gating on/off ratios as well as rectification factors up to ~70 by changing the surface charge of nanofluidic systems. However the use of pH as a stimulus involves changing the electrolyte solutions, and it is difficult to perform reversible on/off experiments. While thus far, uphill active transport has barely been demonstrated using stimuli other than light, the main difficulty that needs to be overcome is the huge sensitivity to small variations in the stimulus (temperature, voltage, ion or pH), which might be difficult to control. Recently, magnetic field control of ionic flow was reported (**Table 2** in the **Supplemental Material**). The basis of this control is the change in the channel entrance size upon application of a magnetic field when the channel is surrounded by ferromagnets bound to the channel entrance. In another mechanism, the channels are initially blocked by a hydrophobic ferromagnetic fluid, and by applying a magnetic field the ferrofluid clears the opening, thus permitting the ionic flow. This magnetic control has been shown to enhance or suppress ionic movement with a very high gating on/off ratio (≈10,000) with excellent stability (up to 130 cycles). Development of neuromorphic devices designed from stimuli-responsive channels is another emerging direction inspired by the energy-efficient computation architectures of biological systems (175). Multistimuli-responsive channels where more than one stimulus is used are highly promising, and prototype nanofluidic logic devices, diodes and gated devices have been demonstrated with this approach (**Table 2** in **supplementary material**).

## 5. Conclusion

In this review, we briefly discuss the roadmap of the fluidics field from micro to nanofluidics and the newly emerging angstrofluidics field, with representative examples of fluidic systems. State-of-the-art angstrochannels with different dimensionalities are discussed, along with advancements in measurement techniques that have been adapted for the ongoing miniaturization. Ionic flow induced by stimuli (electric, pressure, concentration gradient, etc.) and the development of smart devices controlled by other external stimuli (light, pH, temperature, etc.) are reviewed. A number of research groups across the world are working on sub-continuum phenomena with unique architectures for angstrochannels. So far the phenomena probed with angstrochannels have only been a small portion of what is still unknown, and many more discoveries can be expected in this domain. This emerging field will benefit from the new perspective of single-molecule measurement techniques, collaborative efforts combining several different systems, and important theoretical efforts to predict the correct materials and properties to be probed.

---

**Summary points:**

1. Angstrofluidics is an emerging new field made feasible by the availability of new materials that could be used as atomic-scale building blocks of channels. Several angstrochannels with 0D, 1D, and 2D architectures are readily available, and advanced measurement tools are being developed rapidly.

2. Several external stimuli that modify nano- and angstrofluidic ionic transport are reviewed here. The combination of several stimuli for better control of ionic gating, rectification, and active transport through Å-channels will be a next step in the direction of smart ionic devices, iontronics, and neuromorphic devices.

3. Angstrofluidic devices have enormous potential for application in the domains of reverse osmosis, smart biomaterials, molecular filters, biosensors, fluidics-based logic devices, desalination, and bioinspired ionic memory devices.

---

**Future prospects & challenges:**

- Applications of nano- and angstrofluidic systems will be explored further in the coming decade.

- Selectivity in artificial channels is still far from that in biological channels. Greater control over the fabrication and functionalization methods to precisely introduce different charges and chemical moieties on the same channel (mimicking different amino acids in protein channels) could greatly enhance their selectivity and overall output (flow, power, etc.).

- Sub-nanometer channels display discrete quantum effects, molecular ordering, ion pairing, and more; new theoretical descriptions are required to understand the emerging physics behind these effects.

- Mass production and robustness are still the main challenges for large-scale applications of nano- and angstrofluidic devices. The ease and speed of fabrication and reproducibility issues still need to be addressed.

**Disclosure Statement**

The authors are not aware of any biases that might be perceived as affecting the objectivity of this review.

## Acknowledgements

B.R. acknowledges funding from the Royal Society University Research Fellowship (URF\R1\180127) and Enhancement Awards (RGF\EA\181000 and RF\ERE\210016) and funding from the European Union's H2020 Framework Programme/European Research Council Starting Grant (agreement number 852674 – AngstroCAP). A.K. acknowledges Ramsay Memorial Fellowship, and Royal Society International Exchanges grant (IES\R1\201028).

# Supporting information

## Angstrofluidics: walking to the limit

Yi You [1,2,#], Abdulghani Ismail[1,2,#], Gwang-Hyeon Nam[1,2] , Solleti Goutham[1,2], Ashok Keerthi [2,3] and Boya Radha[1,2,*]

[1]Department of Physics and Astronomy, School of Natural Sciences, The University of Manchester, Manchester M13 9PL, United Kingdom

[2]National Graphene Institute, The University of Manchester, Manchester M13 9PL, United Kingdom

[3]Department of Chemistry, School of Natural Sciences, The University of Manchester, Manchester M13 9PL, United Kingdom

[#] these authors contributed equally.

*Correspondence to: radha.boya@manchester.ac.uk

### S1. Ionic Interaction Lengths

Electrostatic lengths relate to ion-ion and ion-surface interactions. At the nano/angstrom scale, these lengths play a major role in the electrokinetic phenomena (briefed below). These lengths are interlinked with each other, and incorporate interactions with the solvent molecules.

- Debye length ($\lambda_D$) is the length after which a charged species or surface electric field is completely shielded. In case of a charged interface, it represents the thickness of the diffuse layer where counter ions are more than co-ions, hence the electro-neutrality is not conserved. The main factor controlling this length is the ionic force of the electrolyte. It is worth mentioning that the charge of the solid interface is not a factor controlling the $\lambda_D$. Typical values of $\lambda_D$ range from few angstroms (Å) to tens of nanometers as the salt concentration is varied from mM to few M (1, 2).

- The Bjerrum length ($l_B$) is the length at which the bare Coulomb energy between two elementary charges is balanced by the thermal fluctuation energy ($k_B T$) (3). Therefore, the principal factors controlling this length are the temperature, the dielectric constant of the interacting medium and the charge on the ions. The typical order of magnitude of this length for monovalent ions in bulk water is 0.7 nm.

- The Gouy-Chapman length (($l_{GC}$) is the distance that a charged species must travel from a charged surface to decrease its electrostatic energy by $k_B T$. At this distance, the electrostatic interaction of the ion with the wall matches the thermal energy. In contrast to the $\lambda_D$, the ($l_{GC}$ depends on the surface charge and is independent of the ionic force of the solution. The order of magnitude of this length depends mainly on the material of the channels and their methods of fabrication as well as the pH of the solution. The $l_{GC}$ is usually smaller than the $\lambda_D$ and ranges from few Å to ~1 nm depending on the surface charge.

- The Dukhin length ($l_{Du}$) is the ratio of surface conductance to the bulk conductance. It is a principal factor affecting ion selectivity and current rectification in nanopores with a maximum rectification occurring when the $l_{Du}$ is similar to the channels' confining dimension (4). The $l_{Du}$ varies from several nanometers to micrometers depending on numerous factors including the electrolyte concentration as well as those that affect $\lambda_D$ and $l_B$ like surface charge and geometry.

**Electrokinetic phenomena:**

- Electroosmosis is the movement of the neutral solvent molecules induced by the electric field applied parallel to a charged surface. The counter-ions, next to the inner immobile Helmholtz layer on the charged surface, migrate and drag the solvent molecules, thus giving rise to plug like electro-osmotic flow.
- Electrophoresis is the movement of charged particles relative to a stationary fluid. If the charged particles have different charge-to-mass (q/m) ratios, it leads to the separation of particles into distinct groups as a function of q/m ratio.
- Ionic streaming: Application of pressure difference across the two sides of a nanochannel induces either streaming potential and/or streaming current. The streaming potential occurs due to the difference in the accumulation of ions on both sides of the channel, which results from ion selectivity. Physically it is the opposite of electroosmosis.

**Table S1**: Summary of ions' key properties.

| Ion | Ionic diameter (Å) | Hydrated diameter (Å) | Hydration free energy (KJ/mol) |
|---|---|---|---|
| $F^-$ | 2.3 (5) | 7.0 (5) | −580 (6) |
| $Cl^-$ | 3.6 (7) | 6.6 (5) | −371 (8) |
| $I^-$ | 4.3 (9) | 6.0 (9) | −254 (6) |
| $ClO_4^-$ | 4.5 (10) | 6.8 (10) | −238 (11) |
| $K^+$ | 3.0 (5) | 6.6 (5) | −351 (12) |
| $Cs^+$ | 3.6 (13) | 6.6 (8) | −376 (14) |
| $Na^+$ | 2.3 (5) | 7.2 (15) | −435 (12) |
| $Li^+$ | 1.9 (5) | 7.6 (5) | −544 (12) |
| $Cu^{2+}$ | 1.5 (16) | 8.4 (17) | −2160 (12) |
| $Ca^{2+}$ | 2.5 (7) | 8.4 (18) | −1650 (12) |
| $Fe^{2+}$ | 1.6 (16) | 8.6 (19) | −1980 (12) |
| $Mg^{2+}$ | 1.4 (5) | 8.6 (5) | −1922 (8) |
| $Ni^{2+}$ | 1.4 (20) | 8.1 (8) | −2106 (8) |
| $Zn^{2+}$ | 1.5 (5) | 8.6 (5) | −2197 (21) |
| $Cd^{2+}$ | 1.9 (5) | 8.5 (5) | −1979 (21) |
| $Al^{3+}$ | 1.1 (16) | 9.6 (5) | −4750 (12) |
| $Fe^{3+}$ | 1.3 (16) | 9.6 (19) | −4365 (22) |

**Table S2**: Various external stimuli that control ion transport under confinement. Abbreviations: AAO: Anodic aluminium oxide, ADA: Adamantanamine hydrochloride, APTMS: (3-Aminopropyl)trimethoxysilane, ATP: adenosine triphosphate, β-CD Beta cyclodextrin, BCP: Block copolymer, BTNI: naphthalene derivative, cAMP: Cyclic 3′,5′-adenosine monophosphate, CB: Cucurbit[8]uril, CNN: Carbon nitride nanotube, co-AAC: co -acrylic acid, $\Delta C$: Concentration gradient, EDA: ethylenediamine, GO: Graphene oxide, HPTS: 8–hydroxypyrene-1,3,6-trisulfonate, , I: Current, $l$: length, MGCB: malachite green carbinol base, N3: cis-bis-(4,4′-dicarboxy-2,2′-bipyridine) dithiocyanato ruthenium(II), PANI: polyaniline, PC: Polycarbonate, PDMAEMA: poly(N,N-dimethylaminoethylmethacrylate), PDMS: Polydimethylsiloxane, PET: Polyethylene terephthalate, PI: Polyimide, PLL: Poly-L-lysine, PM: porous membrane, PNIPAm: Poly(N-isopropyl acrylamidePPy: Polypyrrole, PS-b-PDMAEMA: polystyrene-block-poly(N,N-dimethylaminoethylmethacrylate), P-SPMA: Poly-spiropyran-linked methacrylate, SEBS: wax-elastic copolymer (polystyrene- (ethylene-butylene)-polystyrene and $\Delta T$: Temperature gradient.

| | Stimuli | Material | Dimensionality | Dimensions | Stimuli receptor | Mechanism | Major application(s) | Year | Ref |
|---|---|---|---|---|---|---|---|---|---|
| **Light** | UV | PET | 1D | d = 15–27 nm; $l$ = 12 μm | C4–DNA &MGCB | Under UV illumination, MGCB releases $OH^-$ ions which leads to an increase of pH, thus the conformation of the pH sensitive C4-DNA changes and the channel conductivity decreases | Gating, $I_{on}$(dark) / $I_{off}$(UV) ~ 2 | 2012 | (23) |
| | UV | PET | 1D | d = 10 nm; $l$ = 12 μm | Benzoic acid derivative dimers | UV illumination increases the surface charge by enhancing the dissociation of the benzoic acid derivative dimers | Active transport against $\Delta C = 0.75$ | 2016 | (24) |
| | UV-vis | PET | 1D | $d_{tip}$ = 20 nm; $d_{base}$ = 600 nm; $l$ = 12 μm | Pillararene (host) –azobenzene (guest) | Modification of surface charge (alternating positive and negative charge) by threading/dethreading of host-guest i.e., Pillararene-azobenzene receptors | - Inversion of rectification by alternating cation/anion selectivity with light stimuli<br>- Gating, $I_{on}$(vis) / $I_{off}$(UV) ~ 5<br>- Controlled transport of cargo ATP | 2017 | (25) |
| | UV ($\lambda = 365\ nm$), vis ($\lambda = 430\ nm$) | PI | 1D | d = 11.4–13.5 nm; $l$ = 12 μm | Azobenzene & β-CD | Under UV light, the azobenzene is in the E isomer-form, which has weak association with β-CD while visible light transforms it to Z-form. UV-visible illumination causes the reversible photoisomerization of azobenzene derivatives leading to stretch/shrink motion (like a nanomachine) that can entrap/release β -CD to transport it. | Selective transport of β-CD across the membrane | 2018 | (26) |

| Stimuli | | Material | Dimensionality | Dimensions | Stimuli receptor | Mechanism | Major application(s) | Year | Ref |
|---|---|---|---|---|---|---|---|---|---|
| | UV-vis | Alumina | PM | $d_{centre}$ = 8 nm $d_{top-bottom}$ = 35 nm thickness = 98 µm | N3 & Spiropyran | Modification of surface charge by light illumination | - Nanofluidic diode (rectification ratio for dark ~ 3.2, visible ~ 3.6, UV ~ 4.3) - Gating, $I_{on}$(Vis) / $I_{off}$(darkness) ~ 1.4; $I_{on}$(UV) / $I_{off}$(dark) ~ 2.2 | 2018 | (27) |
| | Visible | CNN | PM | d = 30 nm thickness = 60 µm | CNN | Surface charge distribution; separation of electrons and holes gives rise to membrane potential, responsible for ion transport and pumping | Active transport against $\Delta C$ = 5000 | 2019 | (28) |
| | Visible | GO | 2D | h = 0.92 nm; thickness = 5 µm; | GO | When illuminated at an off centre position on the GO membrane, an electric potential difference is built that can drive the transport of ionic species | - Active transport against concentration gradient $\Delta C$ =10 - Gating, ($I_{on}$ ~ 0 pA, $I_{off}$ ~ 13 pA) - Nanofluidic diode: rectification up to $10^4$ under light illumination | 2019 | (29) |
| | Visible ($\lambda$ = 643 nm) | MoS$_2$ | 0D | d = 3 nm ; l = 0.7 nm | MoS$_2$ | Increase in the surface charge that modifies the channel selectivity | - Two times increase in the osmotic energy - Gating, $I_{on}$(laser) / $I_{off}$(dark) ~ 1.6 | 2019 | (30) |
| **Temperature** | 23–40 °C | PI | 1D | d < 5 nm l = 12 µm | PNIPAm brush | Conformational changes lead to pore diameter change by expansion and contraction of the polymer brush | Gating, $I_{on}$(40°C) / $I_{off}$(23°C) ~ 3.5 | 2009 | (31) |
| | 23–60 °C | PC | PM | d = 18 nm; thickness = 6 µm | SEBS | Conformational changes lead to pore diameter change by expansion and contraction of the wax | Gating, $I_{on}$(40°C) / $I_{off}$(36°C) ~ 10 | 2015 | (32) |
| | 23–100 °C $\Delta T$ with steps of 5 °C | Cellulose nanofibres | PM | Spacing = 0.6 nm thickness = 3 mm | Cellulose molecular chains | Thermally generated voltage is enhanced due to effective sodium ion insertion into the charged molecular chains of the cellulosic membrane | Differential thermal voltage generation, useful for conversion of low grade heat-to-electricity | 2019 | (33) |

| | Stimuli | Material | Dimensionality | Dimensions | Stimuli receptor | Mechanism | Major application(s) | Year | Ref |
|---|---|---|---|---|---|---|---|---|---|
| **pH** | 25 & 50 °C | GO grafted with PNIPAm | 2D | $h$ = 1.8–6 nm thickness = 0.45–1.29 µm | PNIPAm | Variation in temperature changes the PNIPAM interspacing (close/open) depending on PNIPAM grafting density | Gating, $I_{on}$(25°C) / $I_{off}$(50°C) ~ 6 | 2019 | (34) |
| | 4.5 & 8.5 | PET grafted with DNA motor | 1D | $d_{tip}$ = 5–45 nm; $d_{base}$ = 900 nm $l$ = 12 µm | DNA | At pH 4.5, DNA folds into a densely packed structure deceasing the pore size, whereas at pH 8.5 DNA relaxes enhancing the channel conductivity. | Gating, $I_{on}$(pH=8.5) / $I_{off}$(pH=4.5) ~ 10 | 2008 | (35) |
| | 2.8 & 10 | PET | 1D | $d_{tip}$ = 20–80 nm; $d_{centre}$ = 150–300 nm; $l$ = 10.8–11.2 µm | Channels asymmetrically functionalized with polyvinylpyridine and poly(acrylic acid) | Double gates (both ends of the channel) could be opened and closed alternatively/simultaneously under symmetric/asymmetric pH | - Active transport against $\Delta C$ = 10 (asymmetric pH) - Gating (symmetric pH), $I_{on}$(pH=10) / $I_{off}$(pH=2.8) ~ 22 | 2013 | (36) |
| | 3 –11 | AAO | PM | $d$ = 20 nm thickness = 85 µm | Functional groups of alumina & patterned APTMS | Asymmetric charge distribution arising from selective modification | Nanofluidic diode– (Increase of ionic current rectification ratio from ~ 2 (pH=2) to ~ 12  (pH=10.3) | 2013 | (37) |
| | 2.8 –10 | PET | 1D | $d_{tip}$ = 40 nm; $d_{base}$ = 240 nm; $l$ = 12 µm | PS-b-P4VP copolymers & homopolystyrene | Change in the surface charge and wettability | Gating, $I_{on}$(pH=2.8) / $I_{off}$(pH=10) ~ 67 | 2021 | (38) |
| **Magnet** | B = 0.11–0.23 T | PET | 1D | $d_{tip}$ = 210 nm; $d_{base}$ = 2.5 µm; $l$ = 12 µm | Iron powders & PDMS deposited around the PET aperture | Conformational change of the aperture diameter by magnetic field | Gating, $I_{on}$(0.23T) / $I_{off}$(0T) ~ 16 | 2018 | (39) |
| | B = 0.45 T | AAO | PM | $d$ = 45 nm thickness = 0.2 mm | EMG 901 ferrofluid | Opening/blocking of the superhydrophilic channels by the hydrophobic ferrofluid with magnetic field | Gating, $I_{on}$(0.45T) / $I_{off}$(0T) ~ 10000, stability 130 cycles | 2019 | (40) |

| | Stimuli | Material | Dimensionality | Dimensions | Stimuli receptor | Mechanism | Major application(s) | Year | Ref |
|---|---|---|---|---|---|---|---|---|---|
| | $B = 0$–$0.52$ T | AAO | PM | $d = 309$ nm | EMG 901 ferrofluid | Opening/blocking of the superhydrophilic channels by the hydrophobic ferrofluid with magnetic field | - Multilevel ultrafast-response (< 0.1 s) - molecular transport with speed control<br>- Gating, $I_{on}$(0T) / $I_{off}$ (lateral magnetic field) ~ 26000, stability 1h cycling | 2020 | (41) |
| **Molecules** | ATP | PET | PM | $d = 270$-$650$ nm<br>thickness = 12 µm | 3D cross linked DNA (capture probe & linker containing ATP aptamer) | Addition of 3D Y-DNA renders the channels inaccessible to ions by closing them. Upon adding ATP, it competes with the 3D Y-DNA and binds to the capture probe & linker modified channel. This leads to disassembly of the 3D DNA nanostructures from the channel, reopening the pathway for ion conduction through the nanopore | - Gating:<br>$I_{on}$(ATP) / $I_{off}$(Y-DNA) ~ $10^3$-$10^5$ | 2015 | (42) |
| | ADA & $K^+$ | PET | 1D | $d_{tip} = 31$ nm;<br>$d_{base} = 186$ nm<br>$l = 12$ µm | Mono- and multilayers of host: CB guest: BTNI, via layer-by-layer assembly | Multiple gating states enabled from layer-by-layer assembly of supramolecular host – guest (CB and BTNI) on the channel surface. The ADA and $K^+$ dissociate CB+BTNI from the channel surface, thus reversibly gating the channel. | - Bidirectional nanofluidic diode based on inversion and tuning of current rectification, from 0.2 to ~6.5<br>- Gating, $I_{on}$(1st state BTNI+CB) / $I_{off}$(10th state BTNI+CB) $\approx 20$ | 2016 | (43) |
| | cAMP | AAO | PM | $d = 80$ nm;<br>thickness = 86 nm | PNIPAm & ATGPA (cAMP recognition unit) | cAMP binds to ATGPA and disrupts its hydrogen bonds with the adjacent NIPAm units. This makes the PNIPAm stretch freely into swollen state, thus decreasing the current upon cAMP addition. | Gating, $I_{on}$(cAMP) / $I_{off}$(control) ~ 1.25 | 2019 | (44) |
| **Electric** | ±5 V | PET | 1D | $d_{tip} = 4$–$30$ nm;<br>$d_{base} = 500$ nm;<br>$l = 12$ µm | (trimethylsilyl) diazomethane | Nanochannels undergo wetting and dewetting by applying different potentials, due to the water condensation and evaporation, which leads to conducting and non-conducting states | Hydrophobic valve<br>non-conducting state ($|E| \leq 1\,V$)<br>conducting state $|E| \geq 1\,V$ | 2011 | (45) |

| | Stimuli | Material | Dimensionality | Dimensions | Stimuli receptor | Mechanism | Major application(s) | Year | Ref |
|---|---|---|---|---|---|---|---|---|---|
| | ±0.8 V | AAO | PM | d = 28 nm | PPy | The increase in potential changes the PPy redox state and increases the positive charge density on PPy. | - Nanofluidic diode rectification factor increases from 1.6 (E=-0.8 V) to 5.4 (E=0.8 V)<br>- Gating, $I_{on}(0.8V) / I_{off}(-0.8V) \sim 4.6$ | 2017 | (46) |
| | ±1 V | MXene | PM | Spacing ~ 0.7 nm | MXene | The gated voltage changes the surface charge of MXene, thus increasing or decreasing the counter ion interaction with the channel walls. | Gating, $I_{on}(+1V) / I_{off}(-1V) \sim 10$ | 2019 | (47) |
| | ±1 V | AAO/PPy | PM | PPy Channels with d < 0.7 nm thickness = 900 nm | PPy | The applied voltage modulates the surface charge by modifying PPy redox states | - Nanofluidic logic gates<br>- In double gated device $I_{on}(-1V) / I_{off}(+1V) \sim 5.7$<br>- Bidirectional nanofluidic diode based on inversion of current rectification<br>- Double gated active ion transport against $\Delta C = 100$ | 2021 | (48) |
| **Multi stimuli** | T = 25–40 °C pH = 3.6–9.4 | PI | 1D | $d_{tip}$ = 25 nm; $d_{base}$ = 1.2-1.5 µm $l$ = 12 µm | PNIPAm-co-AAc brushes | - Temperature changes trigger the conformational change of the PNIPAm-co-AAc brushes. At high temperature (~40 °C), the polymer brushes collapse, creating an increased pore size thus a high conducting state.<br>- pH variation changes the surface charge of the polymer brushes. | - Gating, $I_{on}(40°C) / I_{off}(25°C) \sim 3$<br>- Nanofluidic diode, increase in rectification factor from ~1 (pH=3.6) to ~ 3 (pH 9.4) | 2010 | (49) |
| | UV pH = 3.5–8 | PET | 1D | $d_{tip}$ = 10 nm; $d_{base}$ = 640 nm; $l$ = 12 µm | HPTS & EDA | - UV: opens the channel by changing the surface charge (deprotonate HPTS)<br>- pH variation changes the surface charge | - Gating, $I_{on}(UV) / I_{off}(dark) \sim 3$<br>- Nanofluidic diode, increase in rectification ratio from ~ 0.7 (pH=3.5, dark) to ~ 18 (pH=8, dark), until ~ 30 (pH=8, UV) | 2014 | (50) |

| | Stimuli | Material | Dimensionality | Dimensions | Stimuli receptor | Mechanism | Major application(s) | Year | Ref |
|---|---|---|---|---|---|---|---|---|---|
| **Multi stimuli** | E = ± 4V<br>pH = 5.5 & 8 | PET | 1D | $d$ = 1–20 nm;<br>$l$ = 12 μm | Negative DNA with/without protonable DNA oligomers | - Voltage: Applying voltage causes the movement of negative channel-attached DNA molecules through the pore thus modifying its aperture (steric current reduction) and conductivity (electrostatic cation mobility enhancement)<br>- pH: Lowering the pH to 5.5 makes the positively charged DNA oligomers to bind to the surrounding negatively charged DNA strands, which creates an electrostatic mesh that closes the pore. At neutral pH, the pore keeps open. | Gating, $I_{on}$(pH=8) / $I_{off}$(pH=5.5) ~ 350 | 2014 | (51) |
| | pH: 2-12<br>E: (-)10–(+)750 mV | PC | 1D | $d_{tip}$ = 85-90 nm;<br>$d_{base}$ = 4 μm;<br>$l$ = 30 μm | PANI | Voltage and pH: modification of surface charge. The charge of polyaniline increases with decreasing pH and increasing potential (3 different states). | - pH: nanofluidic diode rectification factor increases from 1.5 (pH=12) to 52 (pH=2)<br>- Voltage E: nanofluidic diode rectification factor increases from 3 (E<200 mV) and 7.5 (200<E<700 mV) to 17 (E>700 mV) | 2015 | (52) |
| | T.= 25–65°C<br>pH = 2-9 | Heterogeneous membrane: BCP (PS-b-PDMAEMA) + PET | PM BCP: 2D + PET:1D | - BCP: channel height 18 nm;<br>$l$ = 600 nm;<br>- PET: $d_{tip}$ = 15 nm;<br>$d_{base}$ = 0.5 μm<br>$l$ = 12 μm | PDMAEMA | - Temperature: induces conformational changes of the PDMAEMA which changes the pore opening dimensions<br>- pH: variation changes the surface charge, wettability and conformation of the channel thus changing the channel ionic conductivity | - Temperature: nanofluidic diode (increase in rectification factor from ~ 40 (25 °C) to ~ 80 (65 °C)<br>- pH: smart gating $I_{on}$(pH=9) / $I_{off}$(pH=2) ~40<br>- Simultaneous pH and thermal, $I_{on}$(pH=9, 65°C) / $I_{off}$(pH=2, 25°C) ~ 80 | 2016 | (53) |

| | Stimuli | Material | Dimensionality | Dimensions | Stimuli receptor | Mechanism | Major application(s) | Year | Ref |
|---|---|---|---|---|---|---|---|---|---|
| **Multi stimuli** | UV ($\lambda = 365\ nm$)-vis ($\lambda = 430\ nm$)-E: $\pm 5V$ | PI | 1D | $d = \sim 500$ nm; $l = 12$ µm | Azobenzene-derivatives & β-CD | - Light: changes the channel conductance by controlling the surface hydrophilicity. For instance, depending on the nature of the irradiating light (UV or visible), the conformation of azobenzene changes thus leading to its association (or dissociation) with amphiphilic β-CD<br>- Voltage: high voltage switches the hydrophobic azobenzene-decorated pores to hydrophilic, rendering them non-conductive. This process is reversible. | - Light: Switching between non-conducting (UV or visible without β-CD) and conducting states (only β-CD +Visible)<br>- Voltage E: Switching between non-conducting ($E > -2.6\ V$) and conducting states ($E < -2.6\ V$) | 2018 | (54) |
| | pH = 2–11 UV-Vis | PI | 1D | $d_{tip} = 20$ nm; $d_{base} = 400$ nm; $l = 12$ µm | Azobenzene (guest) & β-CD (host) | - Light: visible light irradiation changes the conformation of the guest (cis to trans), enabling its interaction with the host, and increases the ion flow. UV irradiation inverses this and promotes the mismatch of host-guest<br>- pH: variation of pH changes the surface charge, thus tuning the rectification properties of the channel. | - Bidirectional nanofluidic diode based on inversion and tuning of current rectification (up to RF>100)<br>- Gating, I$_{on}$(vis) / I$_{off}$(UV) ~ 100 | 2018 | (55) |
| | T. = 20-70 °C pH = 3–11.5 | PET | 1D | $d_{tip} = 40$ nm; $d_{base} = 2$ µm; $l = 12$ µm | PLL | - Temperature: conformational change at high pH (11.5) between α-helix hydrophilic PLL (conducting state) & β-sheet hydrophobic PLL (non-conducting state)<br>- pH: variation changes the surface charge | - Temperature: gating, I$_{on}$(20°C) / I$_{off}$(70°C) > 100<br>- pH: nanofluidic diode (inversion of rectification factor from ~ 8 (at pH=11.5) to ~ −52 (at pH=3) | 2020 | (56) |

| Stimuli | | Material | Dimensionality | Dimensions | Stimuli receptor | Mechanism | Major application(s) | Year | Ref |
|---|---|---|---|---|---|---|---|---|---|
| pH = 3-11 UV-Vis | | Glass | 1D | d = 200 nm; $l$ = 80 µm | P-SPMA | - Light: UV light transforms the hydrophobic spiropyran groups to hydrophilic merocyanin form, thus switching the channel from the "off" state to the "on" state. The inverse happens with visible light illumination.<br>- pH: variation of pH while UV irradiation, changes the surface charge which leads to variation of the rectification. Without UV irradiation, very high or very low (pH=3 or 11) lead to the change of spiropyran to merocyanine and thus the hydrophobic/hydrophilic character and the conductivity of the channel. | - UV: gating, $I_{on}$(UV) / $I_{off}$(vis) ratio > 30<br>- pH: nanofluidic diode, inversion of rectification; rectification factor varies from ~ 30 (pH=4) to ~ -36 (pH=10)) | 2021 | (57) |

**Supporting Information References**